\documentclass[12pt,aps,pra,groupedaddress,amssymb,amsfonts,nofootinbib,tightenlines]{revtex4}
\usepackage{times,color}

\newcommand{\ignore}[1]{}
\newcommand{\todo}[1]{}
\newcommand{\ftodo}[1]{}
\newcommand{\mComment}[1]{}
\newcommand{\hComment}[1]{}
\newcommand{\gComment}[1]{}
\newcommand{\lComment}[1]{}
\ignore{

pdflatex '\def\scmts{1}\input{cohere}'

}
\ignore{

ptexc cohere

}
\ignore{
\ifthenelse{\isundefined{\scmts}}{}{
\renewcommand{\mComment}[1]{\textcolor{blue}{Manny: #1}}
\renewcommand{\hComment}[1]{\textcolor{red}{Howard: #1}}
\renewcommand{\gComment}[1]{\textcolor{magenta}{Gerardo: #1}}
\renewcommand{\lComment}[1]{\textcolor{green}{Lorenza: #1}}
\renewcommand{\todo}[1]{\noindent Todos: \begin{tabular}[t]{lp{5.5in}}#1\end{tabular}}
\renewcommand{\ftodo}[1]{\noindent Further work needed:\par\noindent\iboxlike{Todos:} \begin{tabular}[t]{lp{5.5in}}#1\end{tabular}}
}
}

\usepackage{calc,ifthen}

\newlength{\elimdepthdim}
\newlength{\elimheightdim}
\newlength{\elimwidthdim}

\newlength{\strutdepthdim}
\newlength{\strutheightdim}
\newlength{\strutwidthdim}
\newcommand{\strutlike}[1]{
\setlength{\strutdepthdim}{\depthof{#1}}
\setlength{\strutheightdim}{\heightof{#1}}
\rule[-\strutdepthdim]{0in}{\strutheightdim+\strutdepthdim}
}
\newcommand{\iboxlike}[1]{%
\setlength{\strutdepthdim}{\depthof{#1}}%
\setlength{\strutheightdim}{\heightof{#1}}%
\setlength{\strutwidthdim}{\widthof{#1}}%
\rule[-\strutdepthdim]{0in}{\strutheightdim+\strutdepthdim}%
\rule{\strutwidthdim}{0in}%
}

\newcommand{\fhf}{\hspace*{\fill}}

\newcommand{\ket}[1]{\qvbar{#1}\qrangle}
\newcommand{\bra}[1]{\qlangle{#1}\qvbar}

\newcommand{\ketbra}[2]{\ket{#1}\bra{#2}}

\newcommand{\cB}{{\cal B}}

\newcommand{\cH}{{\cal H}}

\newcommand{\cP}{{\cal P}}

\renewcommand{\tensor}{\otimes}
\newcommand{\trace}{\mbox{tr}}

\def\id{{\mathchoice {\rm 1\mskip-4mu l} {\rm 1\mskip-4mu l} {\rm
1\mskip-4.5mu l} {\rm 1\mskip-5mu l}}}
\newcommand{\suchthat}{{\;\vrulelike{\{}{1.1pt}\;}}

\newcommand{\rls}{\mathbb{R}}
\newcommand{\cmplx}{\mathbb{C}}

\newcommand{\real}{\mathrm{Re}}

\newcommand{\lie}[1]{\mathfrak{#1}}

\newcommand{\slc}[1]{\mathfrak{sl}_{#1}\cmplx}

\newcommand{\cstar}{\dagger}
\newcommand{\scstar}[1]{{#1}^{\!\cstar}}
\newcommand{\trless}[1]{{#1}_{\!\circ}}

\newcommand{\restrict}{{\upharpoonright}}

\newtheorem{Theorem}{Theorem}

\newtheorem{Problem}[Theorem]{Problem}

\newtheorem{Lemma}[Theorem]{Lemma}
\newtheorem{Corollary}[Theorem]{Corollary}
\newcommand{\proof}{\paragraph*{Proof.}}
\newcommand{\proofof}[1]{\paragraph*{Proof of Theorem~#1.}}
\newcommand{\qed}{\hspace*{\fill}\rule{2.5mm}{2.5mm}%
\vspace*{8pt}\par}

\newcounter{herefignum}

\newcommand{\itb}{{$\bullet$}}

\renewcommand{\iboxlike}[1]{\mbox{\ }}
\renewcommand{\strutlike}[1]{}
\renewcommand{\suchthat}{{\,|\,}}

\def\proof{{\em Proof:  }}

\def\openone{\leavevmode\hbox{\small1\kern-3.8pt\normalsize1}}
\def\RR{{\rm I\kern-.2emR}}

\def\proof{{\em Proof:  }}

\def\openone{\leavevmode\hbox{\small1\kern-3.8pt\normalsize1}}
\def\RR{{\rm I\kern-.2emR}}

\def\id{I}

\renewcommand{\ket}[1]{|{#1}\rangle}
\renewcommand{\bra}[1]{\langle{#1}|}

\newcommand{\bitem}{\begin{itemize}}
\newcommand{\eitem}{\end{itemize}}
\newcommand{\benum}{\begin{enumerate}}
\newcommand{\eenum}{\end{enumerate}}
\newcommand{\beq}{\begin{equation}}
\newcommand{\eeq}{\end{equation}}
\newcommand{\beqa}{\begin{eqnarray}}
\newcommand{\eeqa}{\end{eqnarray}}
\newtheorem{definition}{Definition}
\newtheorem{proposition}{Proposition}
\newcommand{\eproof}{\end{proof}}
\newcommand{\bprop}{\begin{proposition}}

\newcommand{\bdef}{\begin{definition}}

\begin{document}

\title{Generalizations of entanglement based on coherent states and
convex sets}

\author{Howard Barnum}
\email[]{barnum@lanl.gov}
\author{Emanuel Knill}
\email[]{knill@lanl.gov}
\author{Gerardo Ortiz}
\email[]{g\underline{\ }ortiz@lanl.gov}
\author{Lorenza Viola}
\email[]{lviola@lanl.gov}
\affiliation{Los Alamos National Laboratory, MS B256,
Los Alamos, NM 87545} 

\date{\today}

\begin{abstract}
Unentangled pure states on a bipartite system are exactly the coherent
states with respect to the group of local transformations. What
aspects of the study of entanglement are applicable to generalized
coherent states?  Conversely, what can be learned about entanglement
from the well-studied theory of coherent states?  With these questions
in mind, we characterize unentangled pure states as extremal states
when considered as linear functionals on the local Lie algebra. As a
result, a relativized notion of purity emerges, showing that there is
a close relationship between purity, coherence and
(non-)entanglement. To a large extent, these concepts can be defined
and studied in the even more general setting of convex cones of
states. Based on the idea that entanglement is relative, we suggest
considering these notions in the context of partially ordered families
of Lie algebras or convex cones, such as those that arise naturally
for multipartite systems.  The study of entanglement includes notions
of local operations and, for information-theoretic purposes,
entanglement measures and ways of scaling systems to enable asymptotic
developments. We propose ways in which these may be generalized to the
Lie-algebraic setting, and to a lesser extent to the convex-cones
setting. One of our motivations for this program is to
understand the role of entanglement-like concepts in condensed
matter. We discuss how our work provides tools for analyzing the
correlations involved in quantum phase transitions and other aspects
of condensed-matter systems.
\end{abstract}


\maketitle

\pagebreak

\renewcommand{\numberline}{}
\tableofcontents

\pagebreak

\todo{
\itb & Proof of the now-no-longer missing implication in the
characterizations of coherent states. That is proof
that coherent states are those with a sufficiently large stabilizing
subalgebra.
\\
\itb & What else can we say about desirable properties for entanglement
measures and the extent to which they are satisfied by our
relative entanglement measures?
\\
\itb & Can we show that the infimums in the definitions of relative
entanglement measures are achieved? It would seem that there is
sufficient compactness in each case.
\\
\itb & There are more statements in the proof of our characterization
theorem for which it would be good to refer to specific pages.
We should also try to pick one good reference and always
refer to that.
}

\ftodo{
\itb & In order to interpret excitations from the meanfield as
noninteracting particles and say something about entanglement between
the particles, it is necessary to have a particle-number preserving
Lie algebra or family of operators that acts sufficiently
non-trivially, for example, irreducibly on each particle-number
sector.  Except for fermions, it is not obvious how to fit this into
the general semsimple Lie algebra setting.  But we can think about
this more.  
}

\hComment{
Why does ``cat'' suggest ``entanglement''?
Cite and discuss Schroedinger's perspective?
Can we get ``citation ranges'' to work?
}

\section{Introduction}

Entangled states are joint states of two or more distinguishable
quantum systems that cannot be expressed as a mixture of products of
states of each system.  Entangled states can exhibit quantum
correlations between the two systems that have no local classical
interpretation.  One of the most important developments in the study
of quantum mechanics was the characterization of these correlations by
Bell~\cite{Bell64a,Bell93a}, whose many experimental
verifications~\cite{Clauser78a,Aspect81a} (see also~\cite{Rowe2001a} and the
references therein) have given further support to the validity of
quantum mechanics.  Entangled states are now widely considered to be
the defining resource of quantum communication, enabling protocols
such as quantum teleportation~\cite{Bennett93a} and leading to great
improvements in the communication efficiency of certain multi-party
tasks~\cite{Raz99a,Buhrman2000a}.  As a result, entanglement is being
actively investigated both from a physical and from an
information-theoretic perspective.

So far, nearly all studies of entanglement involved two or more
distinguishable quantum subsystems. As a result, investigations of
entanglement have focused on understanding how quantum systems are
made up from subsystems and how this differs from classical systems.
However, there are a number of signs that the assumption of
distinguishable quantum subsystems is too narrow to capture all
properties of states that one might like to ascribe to
entanglement. Several
authors~\cite{Grobe94a,Ivanov95a,Schliemann99a,Schliemann2001a,Paskauskas2001a,Zanardi2001a,Zanardi2002a,Shi2002a,Eckert2002a}
have considered entanglement-related notions for bosons and fermions.
For example, consider the state of one photon in two coupled
cavities. Being the state of one particle, there is a tendency to
expect that there is no entanglement, because one particle cannot be
entangled.  On the other hand, each cavity is a quantum system. From
the point of view of these two quantum systems, the state where the
photon is in an equal superposition of being in either cavity can be
represented as $(\ket{10}+\ket{01})/\sqrt{2}$
and is clearly entangled.  Another
example involving photons is provided by optical ``cat
states''~\cite{Yurke86a,Yurke88a}.  In this case, cat states are quantum
superpositions of sufficiently distinct coherent states in a mode. As the name
suggests, such states are thought to involve entanglement. They
certainly have distinctive non-classical behavior, but since they
exist in a single system (the mode) the strict interpretation of
entanglement based on subsystems would indicate that no entanglement
is present.  A third example is that consisting of a number of
fermions in a lattice. The ``simple'' states for such a system are
described by the so-called Slater determinants (see, for example,
\cite{Blaizot86a}, p.~7), which describe the wavefunction of
noninteracting fermions.  Because the fermions in such a wavefunction
are independent, one expects that no entanglement is present in such a
state.  However, from the point of view of the lattice modes, most
Slater determinants exhibit entanglement~\cite{Zanardi2001b}.  The
three examples make it clear that the presence or absence of
entanglement depends on the physically relevant point of view. Here we
propose that this point of view depends on the relationships between
different Lie algebras of observables that determine the dynamics and
our ability to control the system of interest. In particular,
the extent to which entanglement is present depends on the observables
used to measure a system and describe its states.

One of our goals is to show that the relationships between product
states, separable states and entangled states are at least
Lie-algebraic in nature, and to some extent even more general. This
makes it possible to study the salient features of entanglement
without reference to subsystems, using instead whatever Lie algebras
are physically relevant.  For the case of bipartite quantum systems,
the relevant Lie algebra $\lie{h}$ consists of the unilocal operators
(operators of the form $A\tensor \id$ or $\id\tensor B$). To show that
the ideas of entanglement, separability and product states do not
critically depend on the two subsystems, we provide several ways in
which product states can be characterized in terms of $\lie{h}$
alone. All of these ways lead to the same concept for general
semisimple Lie algebras, namely that of generalized coherent
states~\cite{Gilmore75a,Perelomov85a,Zhang90a}. It is therefore
natural to consider product states to be special kinds of coherent
states.  From this perspective, separable states are mixtures of
coherent states, and pure entangled states are incoherent pure states.
Another way to think about these structures is to realize that the
coherent states are exactly those states which are relatively pure,
that is, extremal with respect to the set of expectations of
observables in the Lie algebra.  Thus, pure states are entangled if
they appear to be mixed with respect to the Lie algebra's
expectations. In the case of bipartite quantum systems, this is an
aspect of entanglement that has long been considered a key
nonclassical property of quantum mechanics: Pure entangled states have
mixed reduced density operators whereas, for example, in classical
probability no pure state can have a mixed marginal. See, for
example,~\cite{Giulini96a}, p.~298, \cite{Peres93a}, p.~116
and~\cite{Cohen-Tannoudji77a}, p.~306.

The recognition that incoherence naturally generalizes entanglement
makes explicit the dependence of the notion of entanglement on the
relevant Lie algebra and makes available the tools of the theory of
generalized coherent states~\cite{Perelomov85a,Zhang90a} for
investigating aspects of entanglement.  To extend the power of this
perspective to the information-theoretic applications of entanglement
requires introducing measures of entanglement, generalizing the ways
in which entanglement can be manipulated and providing a means for
using states as a resource.  In bipartite systems, there is an
abundance of measures of entanglement, many of which generalize
naturally. Further measures arise naturally in the general context and
specialize to potentially interesting measures for multipartite
systems that have not yet been considered.  In bipartite systems, a
key role is played by LOCC (local quantum operations and classical
communication) maps. We propose several classes of maps for general
semisimple Lie algebras that, in the case of bipartite systems, are
related to LOCC.  A desirable property of entanglement measures is
that they are monotone non-increasing under LOCC. We can show
monotonicity properties for some classes of maps in the general
setting. To introduce the notion of states as a resource and enable
asymptotic analysis, we consider schemes for associating Lie algebras
with tensor products of systems defined by a given representation of a
semisimple Lie algebra.

For the purpose of determining what are the essential properties of
states needed to study entanglement, we introduce a setting even more
general than Lie algebras. Since the states when viewed as linear
functionals on observables form a convex cone, we generalize the
definitions to the setting where we have two or more convex cones
related by positive maps. The cones represent the family of states as linear
functionals on the Lie algebras. In the case of bipartite systems,
these are the local Lie algebra and the Lie algebra of all operators.
The map relating the two state spaces is the restriction map of linear
functionals. The definitions relating to separability and entanglement
only require this structure. Entanglement measures can also be defined
based only on convexity, and so can various notions of local
maps.

In taking seriously the idea that entanglement is a relative notion,
one finds that in many cases, there are many more than two relevant
Lie algebras. In the bipartite case, we can consider the hierarchy of
algebras consisting of the trivial Lie algebra, the algebra of
operators acting on the first system, that acting on the second
system, the sum of these, and the algebra of all operators.  When
there are more than two systems, the number of different ways of
combining local Lie algebras multiplies.  For photons, there is the
Lie algebra of passive linear operations, of active linear operations,
and that of all linear and nonlinear operations. To these one might
add the Lie algebras acting locally on the modes, etc.  It is in the
increasing amount of information that is available about states as
more operators are added that crucial quantum properties emerge.  We
believe that in studying a given system, it is beneficial to consider
coherence and entanglement properties at multiple levels.

Independently of the work reported here, Klyachko~\cite{Klyachko2002a}
has recently proposed a generalization of entanglement for
representations of semisimple Lie groups. His starting point is an
extremality property that we use as one of the equivalent
characterizations of product (in general, coherent) states.
Klyachko's work is focused on the geometric invariant theory approach
for investigating states with respect to one Lie group of
operators. This approach leads to useful classifications of the orbits
of states under the Lie group's action. In this context, he discusses
how the notions of classical realism that lead to Bell's
inequalities~\cite{Bell64a} generalize to the Lie algebraic
setting. He also introduces notions of maximal entanglement and
another interesting entanglement measure.

In Section~\ref{sect:bipartite}, we introduce the basic notions
required for generalizing separability and entanglement by reviewing
the example of bipartite systems from the point of view of Lie
algebras and coherence. The generalization to semisimple Lie-algebras
is explained in Section~\ref{sect:Lie}, and the extent to which the
generalization depends only on the relationships between convex cones
is discussed in Section~\ref{sect:convex}.  For reference, the
different settings for studying entanglement are compared in
Table~\ref{table:comparison}. The paper concludes with a discussion of
other relevant examples and the potential applications to condensed
matter.  We assume familiarity with the basic concepts of quantum
information and entanglement. A good reference for quantum information
theory is~\cite{Nielsen2001a}. For reviews of entanglement,
see~\cite{Plenio98a,Brukner01a,Lewenstein2000a}. We also use results
from the basic theory of Lie algebras. Details can be found in books
such as~\cite{Humphreys72a,Varadarajan84a,Fulton91a,Knapp96a}. For physically
motivated treatments of Lie algebras,
see~\cite{Gilmore73a,Wybourne74a,Cornwell89a,Fuchs97d,Georgi99a}.
References for convexity and convex cones include~\cite{HHL89a,Barvinok2002a}.

\section{Entanglement for Bipartite Quantum Systems}
\label{sect:bipartite}

The standard setting for studying entanglement involves two (or more)
distinguishable quantum subsystems forming a \emph{bipartite}
system. The properties of entanglement are most salient if the quantum
subsystems are spatially well separated, with communication between
the sites restricted to classical signals subject to speed-of-light
limitations.  Let the state space of two such quantum subsystems be
given by the Hilbert spaces $\cH_{a}$ and $\cH_{b}$ of dimension $N_a$
and $N_b$, respectively.  The joint state space of the bipartite
system is $\cH_{ab}=\cH_a\tensor\cH_{b}$.  All state spaces and
operator algebras are assumed to be finite dimensional. See
Section~\ref{sect:further} for a brief discussion of the need and
possibilities for extensions to infinite dimensional systems. Product
states are pure states of $\cH_{ab}$ of the form
$\ket{\psi}\tensor\ket{\phi}$. Entangled pure states are states of
$\cH_{ab}$ that are not expressible as a product state.  It is
necessary to generalize the state space to mixtures of pure states,
that is probability distributions over pure states.  For this purpose,
one uses density matrices to represent states. A density matrix $\rho$
is pure if $\rho=\ketbra{\psi}{\psi}$ for some $\ket{\psi}$.
Equivalently, it is pure if $\trace(\rho^2)=1$, or if $\rho$ is
extremal in the set of density matrices (see below).  A separable
state is a mixture of product states. Its density matrix is therefore
a \emph{convex combination} of product states, which is a sum of the
form $\sum_k
p_k\ketbra{\psi_k}{\psi_k}\tensor\ketbra{\phi_k}{\phi_k}$, where
$(p_k)_k$ is a probability distribution~\cite{Werner89a}.  We will use
the expressions ``convex combination'' and ``mixture''
interchangeably.  A non-separable state is said to be entangled. It is
worth recalling that separable states can have non-classical
features. For example, see~\cite{Bennett99b,Ollivier2002a}.

\subsection{Characterizing Product States}

In our approach, the key distinction between entangled and separable
states is the difference between the way things look locally and
globally. The local observables are operators of the form $A\tensor
\id$ and $\id\tensor B$. For our purposes, it is convenient to allow
arbitrary operators as observables, not only hermitian ones. Since
non-hermitian operators can be expressed as complex linear
combinations of hermitian operators, expectations of such operators
are readily computed from expectations of hermitian operators. 

If a pure state of the two systems in unentangled, then it is
completely determined by the expectation values of the local
observables.  To specify a pure entangled state requires knowledge of
the correlations, which are expectations of operators of the form
$A\tensor B$. Note that this method for distinguishing between
unentangled and entangled states does not extend to mixtures. A
generic separable state can contain non-trivial correlations. An
example is
$(\ketbra{0}{0}\tensor\ketbra{0}{0}+\ketbra{1}{1}\tensor\ketbra{1}{1})/2$.
Here the two subsystems are classically correlated.  Nevertheless, it
is possible to characterize separability by investigating the
structure of states in terms of their expectations of local versus
global observables.

There are four non-trivial Lie algebras of observables that determine
the structure of the bipartite system.  Let $\lie{h}_a$ ($\lie{h}_b$)
be the Lie algebra of operators of the from $A\tensor\id$ ($\id\tensor
B$) acting on system $a$ ($b$).  We call these the \emph{unilocal}
algebras, because they consist of operators acting on one subsystem
only.  The \emph{local} Lie algebra is given by
$\lie{h}_l=\lie{h}_a+\lie{h}_b$.  Let $\lie{g}$ be the Lie algebra of
all operators on $\cH_{ab}$. As defined, these four Lie algebras are
complex.  However, as families of operators they are
\emph{$\cstar$-closed}, that is, closed under hermitian
conjugation. Let $\real(\lie{h})$ be the set of hermitian operators in
$\lie{h}$. For a hermitian-closed space of operators $\lie{h}$,
$\lie{h}=\real(\lie{h})+i\real(\lie{h})$, where $i=\sqrt{-1}$.  Using
complex Lie algebras simplifies the representation theory and is
useful for defining generalizations of local quantum maps (see
Section~\ref{sect:bi_lqo}).  Although exponentials $e^{A}$ for
non-skew-hermitian operators are not unitary, they can be interpreted
as Lie-algebraically definable operators associated with postselected
outcomes in an implementation of a quantum map.

A simple way of characterizing product states without referring to the
underlying partition into two subsystems can be based on unique ground
states. A \emph{unique ground state} of a hermitian operator is a
unique minimum-eigenvalue eigenstate.  Operators with degenerate
minimum-eigenvalue eigenspaces do not have a unique ground state.  In
general, we call the the minimum-eigenvalue eigenspace of an operator
the \emph{ground space}.

\begin{Theorem}
\label{thm:prod=ue}
$\ket{\psi}\in\cH_{ab}$ is a product state iff it is the unique
ground state of an operator in $\real(\lie{h}_{l})$.
\end{Theorem}

\proof Suppose that $\ket{\psi}$ is the unique ground state of
$H=A\tensor\id+\id\tensor B\in\real(\lie{h}_{l})$.  The ground space
of $H$ is the intersection of the ground spaces
of $A\tensor\id$ and $\id\tensor B$, which are
product subspaces.  Thus, a unique ground state is a product
state. Conversely, let $\ket{\psi}=\ket{\phi_a}\tensor\ket{\phi_b}$.
Choose an operator $A$ ($B$) on $\cH_a$ ($\cH_b$) such that
$\ket{\phi_a}$ ($\ket{\phi_b}$) is the unique ground state of $A$
($B$). Then $\ket{\psi}$ is the unique ground state of
$A\tensor\id+\id\tensor B\in\real(\lie{h}_{l})$.  \qed

We can use Theorem~\ref{thm:prod=ue} to define a generalization of a
product state for any hermitian-closed Lie algebra of operators.  As
we will see in Section~\ref{sect:Lie}, this generalization agrees
with the notion of generalized coherent states.

The distinction between product and entangled states can also be
viewed in terms of purity with respect to the relevant algebra of
operators. It can be seen that product states are exactly the states
whose reduced density matrices on each of the two subsystems are pure.
The two reduced density matrices for a state completely determine the
expectations of the observables in the local Lie algebra.  To prepare
for generalizing these observations, consider states as linear
functionals on the Lie algebras in question.  We define an
\emph{$\lie{h}$-state} to be a linear functional $\lambda$ on the
operators of $\lie{h}$ \emph{induced} by a density matrix $\rho$
according to $\lambda(C)=\trace(\rho C)$.  The set of $\lie{h}$-states
is denoted by $\lie{h}^{+}$.  In the present setting, states are
completely determined by the linear functional on the Lie algebra of
all operators $\lie{g}$ induced by their density matrix.  A
$\lie{g}$-state $\lambda$ can be restricted to each of the Lie
algebras $\lie{h}_a$, $\lie{h}_b$ and $\lie{h}_{l}$.  For example, the
restriction $\lambda\restrict\lie{h}_a$ of $\lambda$ to $\lie{h}_a$
determines the expectations of observables on the first subsystem, and
therefore the reduced density matrix associated with the state.

Consider the set $\lie{h}_{l}^+$ of $\lie{h}_{l}$-states. This set is
closed under \emph{convex} (or \emph{probabilistic}) combination. That
is, if the $\lambda_k$ are $\lie{h}_{l}$-states, then so
is $\sum_k p_k\lambda_k$ for any probability distribution
$(p_k)_k$. By compactness, all states in $\lie{h}_{l}^+$ can be
obtained as convex combinations of \emph{extremal} states (or
\emph{extreme points} of $\lie{h}_{l}^+$).  Extremal states are states
not expressible as a convex combination of other states.
If the only information available about a state are the expectations
of observables in $\lie{h}_l$, then states that induce extremal
expectations, that is, extremal elements of $\lie{h}_{l}^+$, are
those about which there is the least uncertainty.  It therefore makes
sense to call such states \emph{pure}, or \emph{$\lie{h}_l$-pure}, to be
specific.

\begin{Theorem}
\label{thm:pure=ue}
An $\lie{h}_l$-state is pure iff it is induced by a pure product state.
\end{Theorem}

\proof Consider a density matrix $\rho$ inducing the $\lie{h}_l$-state
$\lambda$. The state $\lambda$ is determined by the reduced density
matrices of $\rho$. It is possible to find a probabilistic combination
of pure product states with the same reduced density matrices, which
therefore also induces $\lambda$. This implies that every 
$\lie{h}_l$-state is expressible as a probabilistic combination of $\lie{h}_l$
states induced by pure product states.  Consequently, the pure
$\lie{h}_l$-states are among those induced by pure product
states. Conversely, if $\lambda$ is not pure, then $\lambda$ can be
nontrivially expressed in the form $p\lambda_1+(1-p)\lambda_2$ where
the $\lambda_k$ are $\lie{h}_l$-states. It follows that the two reduced
density matrices that can be deduced from $\lambda$ are not both pure:
They are mixtures of the reduced density matrices deduced from
$\lambda_k$, and since $\lambda_1\not=\lambda_2$, at least one of
these mixtures is nontrivial.  \qed

The previous theorem shows that the difference between pure
unentangled states and pure entangled states is that as expectations
of $\lie{h}_l$, the latter are not extremal. If the only information
that is available are expectations of $C\in\lie{h}_l$, it is not
possible to distinguish between entangled states and unentangled mixed
(that is, separable) states. To distinguish, we need expectations of
other operators.  It is worth noting what it means to have access only
to expectations of sets of observables. Given only a single instance
of a quantum system, the expectations cannot be inferred.  On the
other hand, with sufficiently powerful control, it is possible to
realize a projective measurement of the eigenvalues of observables, a
process that gives information not just about the expectation of an
observable, but also about the expectations of its powers.  One
situation where access to expectations only is realistic is when the
quantum system must be accessed collectively in large ensembles
involving mostly identical states. In an appropriate weak interaction
and large ensemble limit, the effect on other large systems reveals
the expectations of observables involved in the interaction, whereas
the effect on the systems in the ensemble tends to a unitary evolution
with the observable as a Hamiltonian.  The weak interaction therefore
naturally limits the available control to Lie algebras generated by a
small number of observables.  An example where this situation occurs
for systems that are best modeled as being quantum is nuclear magnetic
resonance of molecules in the liquid state~\cite{Abragam61a}.

\subsection{Local Quantum Maps}
\label{sect:bi_lqo}

One can compare states in the context of information processing
resources by considering families of ``local'' quantum maps that can
be used to convert states. For bipartite systems, as well as for
multipartite systems in general, the most important such family, LOCC,
consists of maps that can be implemented with local quantum maps with
access to ancillas and classical communication (see
\cite{Nielsen2001a}, Sect. 12.5). A larger family, the separable
quantum maps, have an operator-sum representation consisting of
operators of the form $A\tensor B$.  Separable quantum maps are
readily generalized to the Lie algebraic setting, whereas we have not
yet found an equally convincing generalization of LOCC.

A \emph{quantum map} is a trace-preserving completely-positive linear
transformation of density operators. Rather than define these terms,
we use the fact that every quantum map can be written in the
operator-sum representation as $\rho\rightarrow\sum_k C_k\rho
\scstar{C_k}$, with $\sum_k \scstar{C_k} C_k = \id$. We will also
consider \emph{completely-positive maps}, which have the same form,
but don't require the constraint on the $C_k$.  To define LOCC, we
make the sequence $\mathbf{C}=(C_k)_k$ explicit and define
$\mathbf{C}(\rho)=\sum_k C_k\rho \scstar{C_k}$.  Note that the
sequence $\mathbf{C}$ is not uniquely determined by the map.  We call
$\mathbf{C}$ an \emph{explicit map}.  See~\cite{Nielsen2001a}, p. 372
for how to determine when two explicit maps act the same.  To avoid
trivial degeneracies, we assume that the operators that define an
explicit map are always non-zero.  If $(\mathbf{D}_k)_k$ is a sequence
of explicit quantum maps, then the conditional composition of
$\mathbf{C}$ and $(\mathbf{D}_k)_k$ is the quantum map with operator
sequence $(D_{kl}C_k)_{kl}$ and action
$\rho\rightarrow\sum_{kl}D_{kl}C_k\rho \scstar{C_k}
\scstar{D_{kl}}$. A \emph{unilocal quantum map} is a map of the form
$(A_k\tensor\id)_k$ or $(\id\tensor B_k)_k$.  LOCC is the set of
quantum maps obtained as conditional compositions of unilocal
maps. The length of the composition is associated with the number of
rounds of classical communication.  A \emph{separable map} is a
completely positive map with an explicit form given by $(A_k\tensor
B_k)_k$. Note that all LOCC maps are necessarily separable.  The set
of separable maps has been called
SLOCC~\cite{Vidal2000b,Bennett2001a,Duer2001a} and can be viewed as
maps that can be implemented with LOCC and postselection based on the
communication record.

Quantum maps as defined here are often called ``quantum
operations''~\cite{Nielsen2001a}, though the latter term is sometimes
extended to include non-trace-preserving completely positive maps.  In
this manuscript, we use the word ``map'' to refer to linear functions
of spaces other than the Hilbert space of the quantum system under
consideration. We use the word ``operator'' to refer to linear
functions from the Hilbert space to itself.  An important role in
defining various notions of local maps is played by explicit maps,
which in the bipartite and in the Lie algebraic setting are
completely-positive by definition.  There is the potential for
confusion in referring to explicit maps.  For example, an explicit map
can be separable without the operators in the explicit representation
having the necessary product form.  To simplify the terminology, we
position the adjective ``explicit'' such that it applies to all
modifiers between it and the word ``map''.  For example, an explicit
separable map $\mathbf{C}=(C_k)_k$ satisfies that each $C_k$ is a
product operator, whereas this is not required of separable explicit
maps.

Separable maps can be defined from $\lie{h}_l$ without
reference to the two component subsystems.
\begin{Theorem}
\label{thm:clieh=sep}
A completely positive map is separable iff it has an explicit
representation $(C_k)_k$ with $C_k\in \overline{e^{\lie{h}_l}}$.
\end{Theorem}

By definition, $\overline{e^{\lie{h}_l}}$ is
the topological closure of the set of all exponentials
of operators in $\lie{h}_l$. The notion of closure may
be based on the norm induced by the matrix inner product
$\trace(A^\cstar B)$.

\proof $e^{\lie{h}_l}$ consists of all non-zero determinant operators of
the form $A\tensor B$. Thus $e^{\lie{h}_l}$ contains all
invertible product operators, which are dense in the set of product
operators.  The set of product operators is closed.  \qed

There are separable quantum maps that are not
LOCC~\cite{Bennett99b}. The goal is to define or construct, with
minimal reference to the two subsystems, quantum maps that respect
locality better than separable ones. For example, in order to
construct the family of LOCC maps, it is sufficient to be able
to determine when an operator in $\lie{h}_l$ is unilocal, and when a
family of unilocal operators all act on the same side.  With this
ability, one can construct LOCC as was done above, by conditional
composition. If the ability does not depend on the bipartite nature of
the system, there is hope that LOCC has a non-trivial generalization.

We have two approaches to obtaining families of separable quantum
maps with stronger locality properties.  The first approach is based
on the observation that unilocal maps induce well-defined
transformations of $\lie{h}_a$-, $\lie{h}_b$- and $\lie{h}_l$-states.
To formally define what this means, let $\mathbf{C}$ be an explicit
map. Then $\mathbf{C}$ acts on the set of linear functionals
$\lie{g}^*$ of $\lie{g}$ according to $\mathbf{C}(\lambda)(X) =
\lambda(\sum_k \scstar{C_k} X C_k)$. 
\ignore{

Some comment about the ``Heisenberg'' picture in the argument may
be appropriate here.

}%
It will be clear from context
whether we are applying $\mathbf{C}$ to operators or to linear
functionals.  The map $\mathbf{C}$, but not its explicit form, is
determined by the action on $\lie{g}$-states.  Note also that
$\lie{g}$-states linearly span all linear functionals on $\lie{g}$,
and similarly for $\lie{h}$ states with $\lie{h}$ one of $\lie{h}_a$,
$\lie{h}_b$ or $\lie{h}_l$.  $\mathbf{C}$ induces a well-defined
transformation of $\lie{h}$-states if we can complete the following
commutative diagram with a map $\mathbf{C}'$ of $\lie{h}^*$:
\begin{equation}
\label{eq:lift}
\begin{array}{rcccl}
  &\lie{g}^* & \stackrel{\mathbf{C}}{\longrightarrow} & \lie{g}^* &\\
  \textrm{restrict}& \downarrow & & \downarrow &\textrm{restrict}\\
  &\lie{h}^* &\stackrel{\mathbf{C}'}{\longrightarrow} & \lie{h}^*&
\end{array}
\end{equation}
where $\lie{h}^*$ is the set of linear functionals on $\lie{h}$.
Equivalently, whenever $\lambda_1$ and $\lambda_2$ are
$\lie{g}$-states that agree on $\lie{h}$, that is, for which
$\lambda_1\restrict\lie{h}=\lambda_2\restrict\lie{h}$, it is the case that
$\mathbf{C}(\lambda_1)\restrict\lie{h}=
\mathbf{C}(\lambda_2)\restrict\lie{h}$.  Equivalently, if $\lambda$ is
a linear functional on $\lie{g}$ such that
$\lambda\restrict\lie{h}=0$, then
$\mathbf{C}(\lambda)\restrict\lie{h}=0$.  The last statement is equivalent
to the statement
that $\mathbf{C}$ preserves the nullspace of the restriction map.  If any
of the above properties hold, we say that $\mathbf{C}$ can be
\emph{lifted} to $\lie{h}$. Its \emph{lifting} is the map
$\mathbf{C}'$ induced on $\lie{h}$-states.

In the present setting, the notion of liftability can be simplified by
using the canonical (via the trace inner product) isomorphism $\mu$
between $\lie{h}^*$ and $\lie{h}$.  Because the trace inner product is
non-degenerate when restricted to the $\cstar$-closed set of operators
$\lie{h}$, the isomorphism $\mu$ is uniquely determined by the
identity $\lambda(C)=\trace(\mu(\lambda)^\cstar C)$ for all
$C\in\lie{h}$.  In particular, for the algebra $\lie{g}$ of all
operators on $\cH_{ab}$, if $\lambda\in\lie{g}^*$ is induced by the
operator $X$, then $\mu(\lambda)=X$. In general, we say that the
linear functional $\lambda$ is \emph{induced} by $\mu(\lambda)$. Let
$\trace_a$ ($\trace_b$) denote the partial trace mapping operators on
$\cH_{ab}$ to operators on $\cH_{b}$ ($\cH_{a}$, respectively).  We
have the following identities:
\begin{eqnarray*}
\mu(\lambda\restrict\lie{h}_a) &=&
   \trace_b(\mu(\lambda)) \tensor \id/N_b, \\
\mu(\lambda\restrict\lie{h}_b) &=&
   \id/N_a\tensor \trace_a(\mu(\lambda)), \\
\mu(\lambda\restrict\lie{h}_l) &=&
   \trace_b(\mu(\lambda))\tensor \id/N_b +
   \id/N_a\tensor \trace_a(\mu(\lambda)) -
    \trace(\mu(\lambda))(\id\tensor\id)/(N_aN_b).
\end{eqnarray*}
These identities witness the fact that the reduced density matrices of
a state determine the induced linear functionals on the local Lie
algebras.  In the range of $\mu$, the nullspaces of the restriction
maps to $\lie{h}_a, \lie{h_b}$, and $\lie{h}_l$ correspond to the
spaces spanned by $A\otimes B$ with $B, A$ and both $A$ and $B$,
respectively, traceless.  Using the fact that product operators are a
basis of all operators on $\cH_a\tensor\cH_b$, it can be seen that the
explicit map $\mathbf{C}$ lifts to $\lie{h}_a$ iff $ \trace_b\sum_k
C_k (A\tensor B) \scstar{C_k} = \mathbf{C}'(A) \trace(B)$ for some map
$\mathbf{C}'$.  \ignore{

It is not obvious that $\mathbf{C}'$ is necessarily a completely-positive
map.

}%
Equivalently, it lifts iff whenever $\trace(B)=0$,
then $\trace_b\sum_k C_k (A\tensor B) \scstar{C_k} = 0$.  Similar
statements can be made about $\lie{h}_b$.  $\mathbf{C}$ lifts to
$\lie{h}_l$ iff whenever both $\trace(A)=0$ and $\trace(B)=0$, then
$\trace_b\sum_k C_k (A\tensor B) \scstar{C_k} = 0$ and $\trace_a\sum_k
C_k (A\tensor B) \scstar{C_k} = 0$.

Most completely positive maps, even LOCC ones, cannot be
lifted. An example for two qubits is the ``conditional reset''
map that first measures qubit $a$, and if the measurement
outcome is $\ket{1}$, it resets qubit $b$ to $\ket{0}$. However, the
unilocal maps are liftable. In fact, they are liftable to both
$\lie{h}_a$ and $\lie{h}_b$, as are (unconditional) compositions
of unilocal maps. This is the case because such maps are
determined by their actions on the reduced density matrices.  This
suggests that liftable explicit quantum maps could be used as a
generating set for quantum maps with more locality then
separable quantum maps. We next discuss some of the properties
of liftable separable maps and their relationship to LOCC.

\begin{Theorem}
\label{thm:liftone_diu}
Let $\mathbf{C}=(C_1\tensor C_2)$ be a one-operator,  
explicit separable map liftable to $\lie{h}_l$.
Then $C=\alpha U\tensor V$ with $U$ and $V$ unitary.
\end{Theorem}

\proof Liftability implies that if $A$ and $B$ are traceless, then
$C_1 A \scstar{C_1}$ and $C_2 B \scstar{C_2}$ are traceless.  This
implies that the map $f:A\rightarrow C_1 A \scstar{C_1}$ satisfies that
$\trace(f(A)) = \alpha_1\trace(A)$ for some $\alpha_1$. Thus
$f/\alpha_1$ is trace preserving, from which it follows that
$\scstar{C_1} C_1=\alpha_1\id$.  For the same reason, $\scstar{C_2} C_2
= \alpha_2\id$.  The conclusion of the theorem now follows, with
$\alpha = \sqrt{|\alpha_1\alpha_2|}$. \qed

\begin{Theorem}
\label{thm:liftid_uni}
Let $\mathbf{C}$ be an explicit separable map
that lifts to the identity map on $\lie{h}_b$. Then
$\mathbf{C}$ is unilocal, acting on system $a$ only.
\end{Theorem}

\proof Write $\mathbf{C}=(D_k\tensor E_k)_k$, where
$D_k\tensor E_k\not=0$ for all $k$.  By assumption and
applying the map to $\id\tensor B$, 
\begin{equation}
\label{eq:liftid_uni}
\sum_k \trace(D_k \scstar{D_k}) E_k B \scstar{E_k} = N_a B.
\end{equation}
If for some $k$, $E_k\not\propto\id$, we can find
$\ketbra{\psi}{\psi}$ such that $E_k\ketbra{\psi}{\psi}\scstar{E_k}$'s
one-dimensional range does not contain $\ket{\psi}$. Because for all $l$,
$\trace(D_l\scstar{D_l})>0$, the left side of
Equation~\ref{eq:liftid_uni} also has this property, contradicting the
identity.  Hence $E_k = \alpha_k \id$ for each $k$ and the result
follows.  \qed

Theorem~\ref{thm:liftid_uni} characterizes unilocal maps but has the
disadvantage that we have to refer explicitly to the unilocal Lie
algebras, thus requiring more information about locality than that
provided by $\lie{h}_l$ alone. This suggests the following
problem:

\begin{Problem}
\label{prob:sep_lift}
Are separable quantum maps that lift to $\lie{h}_l$
\emph{LOCC}? Are they \emph{LOCC} if they lift to both $\lie{h}_a$ and
$\lie{h}_b$?
\end{Problem}

If the answer to this problem is ``no'', then we are interested in the
question of whether the explicit separable quantum maps that are
liftable to $\lie{h}_l$ generate all separable quantum maps by
conditional composition.

In order to be able to conditionally compose  explicit separable 
quantum maps that are LOCC without departing from LOCC, we need
the explicit representations to have the additional property that they
can be LOCC implemented in such a way that the communication record
reveals which of the operators in the sequence occurred. 
Following our convention for using the adjective ``explicit'',
we call an explicit quantum map with this property 
an \emph{explicit LOCC map}.

\begin{Problem}
\label{prob:exsep_locc}
Are there explicit separable quantum maps that are \emph{LOCC}
but not explicit \emph{LOCC}?
\end{Problem}

If the answers to this problem and to Problem~\ref{prob:sep_lift} are
``yes'', then one has to consider the strengthening of the questions
in Problem~\ref{prob:sep_lift} where ``separable'' is replaced by
``explict separable'' and ``LOCC'' by ``explicit LOCC''. This is
required so that conditional composition can be used without leaving
LOCC.  Here is one case where we can prove that a family of quantum
maps is explicit LOCC.

\begin{Theorem}
\label{thm:nsep_locc}
Let $\mathbf{C}=(D_k\tensor E_k)$ be an explicit separable quantum map
that lifts to $\lie{h}_l$ with the additional property that
$(\scstar{D_k} D_k)_k$ is linearly independent.  Then $E_k = \gamma_k
U_k$ with $U_k$ unitary.  In particular, $\mathbf{C}$ is an explicit
\emph{LOCC} map.
\end{Theorem}

\proof Using the identification of linear functionals with operators,
consider linear functionals $\lambda$ of $\lie{g}$ induced by
$A\tensor B$ with $\trace(B)=0$.  The restriction of $\lambda$ to
$\lie{h}_l$ is induced by $\trace(A)\id\tensor B\in\lie{h}_l$.  The
restriction has only scalar dependence on $A$.  
Restricting after
applying $\mathbf{C}$ gives the linear functional induced by
\begin{equation}
  \sum_k\trace(D_k A \scstar{D_k}) \id/N_a \tensor E_k B \scstar{E_k} 
    + 
  \sum_k D_k A \scstar{D_k} \tensor \trace(E_k B \scstar{E_k})\id/N_b
\end{equation}
Note that because $\trace(B) = 0$, and the assumption
that the map is trace preserving, the contribution to
$\id\tensor\id$ vanishes.
Because of liftability, the same scalar dependence applies
to this expression. By cyclicity of the trace,
$\trace(D_k A \scstar{D_k}) = \trace(A \scstar{D_k} D_k)$.
Because the $\scstar{D_k}D_k$ are independent,
we can choose
$A_l$ such that $\trace(A_l\scstar{D_k} D_k) = \delta_{lk}$.
Hence the following are all scalar multiples of
the same operator, where the scalar is independent of $B$:
\begin{equation}
O_l=
  \id/N_a \tensor E_l B \scstar{E_l} 
    + 
  \sum_k D_k A_l \scstar{D_k} \tensor \trace(E_k B \scstar{E_k})\id/N_b.
\end{equation}
Computing the partial trace over the first system, we get
\begin{eqnarray*}
\begin{array}{rcll}
\trace_a(O_l) &=&
  E_l B \scstar{E_l}
    + \sum_k \trace(A_l \scstar{D_k} D_k)\trace(B\scstar{E_k}E_k)/N_b
    & \textrm{by cyclicity of trace,}\\
  &=& E_l B\scstar{E_l}
    + \sum_k \trace((A_l\tensor  B)(\scstar{D_k}D_k\tensor\scstar{E_k}E_k))/N_b
     &\textrm{because $\trace$ is multiplicative for $\tensor$,}\\
  &=& E_l B\scstar{E_l}
    + \trace(A_l\tensor B)/N_b & \textrm{because $\mathbf{C}$ is a quantum map,}\\
  &=& E_l B\scstar{E_l} & \textrm{because $\trace(B)=0$.}
\end{array}
\end{eqnarray*}
Consequently, the operators $E_l B \scstar{E_l}$ are all proportional
with constant of proportionality independent of $B$. Consider
$E=E_r$. We have
\begin{equation}
E_l B \scstar{E_l} = \alpha_l E B E^\cstar
\end{equation}
for all traceless $B$, where $\alpha_l\trace(A_r) = \trace(A_l)$.
Reformulating, we get that for all traceless $B$,
$\trace(B \scstar{E_l} E_l) = \alpha_l\trace(B E^\cstar E)$.
Hence $\scstar{E_l} E_l = \alpha_l E^\cstar E + \beta_l\id$ for
some $\beta_l$. The trace-preserving condition
requires that
\begin{eqnarray}
\id\tensor\id &=& \sum_k \scstar{D_k} D_k \tensor \scstar{E_k} E_k \label{eq:tp1}\\
  &=& \sum_k \scstar{D_k} D_k \tensor \left(\alpha_k E^\cstar E + \beta_k\id\right) 
      \nonumber\\
  &=& \left(\sum_k \alpha_k \scstar{D_k} D_k\right)\tensor E^\cstar E +
      \left(\sum_k \beta_k \scstar{D_k} D_k\right)\tensor\id
\end{eqnarray}
Suppose that the traceless part of $E^\cstar E$ is not zero, Then
$\sum_k \alpha_k\scstar{D_k} D_k = 0$, which is possible only if
$\alpha_k=0$ for all $k$ (by independence). But by construction
$\alpha_r=1$, so $E^\cstar E$ is a multiple of the identity, hence
$E=E_r$ is a multiple of a unitary operator, say $E_r=\gamma_r U_r$.
Returning to the trace-preserving condition (Equation~\ref{eq:tp1}) and using
the fact that $r$ was arbitrary, we find that $\sum_k
\gamma_k \scstar{D_k} D_k \bar\gamma_k = \id$.  This makes
$\mathbf{D}=(\gamma_k D_k\tensor \id)_k$ a unilocal quantum map.
The $U_k$ can be implemented conditionally on which $D_k$ occurs in a
unilocal implementation of $\mathbf{D}$, hence $\mathbf{C}$ is LOCC.
\qed

\begin{Corollary}
\label{cor:2sep_locc}
Let $\mathbf{C}=(D_1\tensor E_1,D_2\tensor E_2)$ be an
explicit separable quantum map that lifts to $\lie{h}_l$. 
Then $\mathbf{C}$ is explicit \emph{LOCC}.
\end{Corollary}

\proof The result follows by Theorem~\ref{thm:nsep_locc} unless
$\scstar{D_2} D_2 = \alpha' \scstar{D_1} D_1$ and $\scstar{E_2} E_2 =
\beta' \scstar{E_1} E_1$ for some $\alpha$ and $\beta$. In this case,
using the trace-preserving condition, $\scstar{D_1} D_1\tensor
\scstar{E_1} E_1 \propto \id\tensor\id$ making all operators
proportional to unitaries.  Such an map can be realized
explicitly with LOCC by first creating a shared random variable, then
implementing local unitaries conditional on the random variable.  \qed

Every explicit unilocal quantum map can be obtained as a composition
of binary unilocal quantum maps, where a \emph{binary} quantum map is
an explicit quantum map consisting of two operators. The modifier
``explicit'' is assumed when using the modifier ``binary''. We can therefore
use the corollary to characterize LOCC as the quantum maps obtained by
conditional composition of binary separable quantum maps that
lift to $\lie{h}_l$.

Instead of using liftability as the basis for generalizing LOCC and
other classes of local maps, one can use the spectral properties of the
constituent operators of an explicit quantum map.  This idea is motivated
by the following result:

\begin{Theorem}
\label{thm:maxeig=unil}
An operator in $\real(\lie{h}_l)$ that has a
maximal ground space is unilocal.
\end{Theorem}

Maximal means maximal among ground spaces
different from $\cH$ of operators in $\real({\lie{h}_l})$.

\proof An operator in $\real(\lie{h}_l)$ is of the form
$A\tensor\id+\id\tensor B$. By subtracting a multiple of the identity,
we can assume that $A$ and $B$ traceless, not both zero.  If they are
both non-zero, then the operator's ground space is strictly contained
in that of $A\tensor\id$, hence not maximal.  \qed

For future reference, an operator whose traceless part is zero or
satisfies the condition of Theorem~\ref{thm:maxeig=unil} is said to be
\emph{maximally unilocal}.  Note that except for $N_a=N_b=2$, not all
unilocal operators in $\real(\lie{h}_l)$ are maximally unilocal.
However, two maximally unilocal operators $C_1$ and $C_2$ with ground
spaces $H_1$ and $H_2$ such that $H_2=e^{D}H_1$ for some
$D\in\lie{h}_l$ act on the same side.  Also, if $C_1$ is maximally
unilocal and $C_2=e^{D}C_1e^{-D}$ with $D\in\lie{h}_l$, then $C_2$ is
unilocal and acts on the same side. We call a family of operators
contained in the span of $\{e^{D}Ce^{-D}\suchthat D\in\lie{h}_l\}$
with $C$ maximally unilocal an \emph{m-compatible} unilocal family.
With this definition, we have:

\begin{Theorem}
\label{thm:allunis}
An explicit  unilocal quantum map consists of an m-compatible
unilocal family of operators.
\end{Theorem}

\proof Every unilocal one-dimensional projector is maximally unilocal,
and the span of the conjugates under $e^{\lie{h}_l}$ of one such
projector consists of all operators acting on the same side.  \qed

Using this theorem, we can characterize LOCC as the set of quantum
maps obtained by conditional composition of explicit m-compatible
quantum maps. However, this characterization is not directly related
to the definition of separable maps.  To do so requires introducing
explicit quantum maps whose operators are exponentials of members of
an m-compatible family. Also note that in addition to using linear
closure in the definition of m-compatibility, we could have used
closure under commutators (Lie bracket). In the bipartite setting,
this makes no difference. Alternatively, we could have left out linear
closure and just used conjugation under $e^{\lie{h}_l}$.  We do not
know whether conditional composition of the resulting quantum maps
yields LOCC. See the discussion of this topic in
Section~\ref{sect:li_lqo}.

\subsection{Communication Complexity}

In the study of multiparty protocols, an important issue is the
communication complexity of converting one state to another using LOCC
maps. The communication complexity is defined as the number of
classical bits that need to be communicated.  The communication
complexity of a particular LOCC map to a given state can
be determined from a representation as a conditional composition.
This can be done by adding the resources used in each round.  The
contribution from a round depends on the previous map in the
sequence of conditional compositions, as we now explain.  Suppose that
the initial state's density matrix is given by $\rho$, the total
explicit quantum map before the round under consideration is
$\mathbf{C}$, and this is then conditionally composed with the family
of unilocal explicit quantum maps $\mathbf{B}_k$.  In general,
given an explicit quantum map $\mathbf{D}$ applied to density
matrix $\rho$, the average number of bits needed to communicate the
outcomes is given by $H(\mathbf{D},\rho)=-\sum_k p_k\log p_k$, where
$p_k=\trace(\rho \scstar{D_k} D_k)$ is the probability of outcome
$D_k$. This is of course an asymptotic expression assuming knowledge
of $\rho$. In other cases one might prefer to just use
$\log|\mathbf{D}|$ as the number of bits required.  In any case, the
contribution to the communication complexity of the current round
is the average communication 
complexity for transmitting the information in the outcomes
of the conditionally applied maps. This quantity is given by
\begin{equation}
\label{eq:comcplx}
\sum_k \trace (\rho \scstar{C_k} C_k) H(\mathbf{B}_k,C_k\rho \scstar{C_k}/\trace (\rho \scstar{C_k} C_k)).
\end{equation}
The contributions from each round are added up to obtain the
communication complexity of the sequence of conditional
compositions. Depending on the application, the contribution of the
last round can be omitted as its outcomes need not be communicated to
implemented the quantum map.  Note that if the detailed outcomes in
one round are not required for conditioning in the next rounds, then
the explicit maps can be modified to defer these outcomes until the
last round, which is one reason to omit the contribution of the last
round.

In general, the goal is to implement a given communication task with
(near) minimum communication complexity.  By determining the
complexity according to Expression~\ref{eq:comcplx}, we can generalize
communication complexity to any scheme for defining a family of
quantum maps by conditional composition, including the generalized
local maps to be introduced for the Lie algebraic setting in
Section~\ref{sect:Lie}.

\subsection{Resource Scaling}
\label{sect:bi-resources}

An important aspect of information theory involves asymptotic
characterizations of the relationships between information resources
and of the complexity of tasks. To asymptotically scale up a problem,
one usually creates tensor copies of the bipartite states involved and
then investigates their relationships in the context of the now much
larger bipartite system. The relationship between the local Lie
algebras of the individual bipartite subsystems and the one obtained
after forming the tensor products requires a construction other than
the usual products. We did not find an obvious way of implementing
such a construction that does not rely on knowledge of additional
structure. It may be the case that one must have knowledge of how the
representation of $\lie{h}_l$ was constructed.  Nevertheless, there
are a few things we can say that may help in better understanding how
resources can be scaled and how to implement asymptotic analyses.

We construct the space $\cH=\cH_{ab}\tensor\ldots\tensor\cH_{ab}$ as
an $n$-fold tensor product of copies of $\cH_{ab}$.  Let
$\lie{h}_{l,k}$ be the local Lie algebra acting on the $k$'th
factor. 
Let $\lie{h}_{L}$ be the local Lie algebra for $\cH$, where
$\cH$ is bipartitioned into $\cH_a\tensor\ldots\tensor\cH_a$ and
$\cH_b\tensor\ldots\tensor\cH_b$. Define $\lie{h}_{a,k}$,
$\lie{h}_A$, $\lie{h}_{b,k}$ and $\lie{h}_B$ likewise.  The group of
permutations on $n$ elements acts on $\cH$ by permuting the tensor
factors.  The goal is to establish how $\lie{h}_L$ relates to the
$\lie{h}_{l,k}$.  It suffices to consider the case $n=2$, because we
can view $\lie{h}_{L}$ as the smallest Lie algebra that contains the
appropriate Lie algebras obtained for each pair of factors.

Let $G_2(\lie{h}_x)$ ($x\in\{a,b,l\}$) be the set of operators $C$ on
$\cH_{ab}\tensor\cH_{ab}$ such that for all operators $X$ on
$\cH_{ab}$, $\trace_1(C(X^\cstar\tensor \id))\in\lie{h}_{x,2}$ and
$\trace_2(C(\id\tensor X^\cstar))\in\lie{h}_{x,1}$.  Here, $\trace_i$ is
tracing out the $i$'th factor with respect to the tensor product
$\cH_{ab}\tensor\cH_{ab}$.  In words, $G_2(\lie{h}_x)$ is the set of
operators which look locally like operators in $\lie{h}_x$.

\begin{Theorem}
\label{thm:tup_trace}
$G_2(\lie{h}_a) = \lie{h}_A$, $G_2(\lie{h}_b)=\lie{h}_B$,
but $G_2(\lie{h}_l)$ strictly contains $\lie{h}_L$.
\end{Theorem}

\proof The definition ensures that $\lie{h}_A\subseteq
G_2(\lie{h}_a)$.  Let $C\in G_2(\lie{h}_a)$.  We can write
$C=\sum_{klrs} \alpha_{klrs} (A_k\tensor B_l)\tensor(A_r\tensor B_s)$
with $(A_k)_k$ and $(B_l)_l$ orthonormal bases of operators including
the identity.  The ordering of the tensor product is according to
$(\cH_a\tensor\cH_b)\tensor(\cH_a\tensor\cH_b)$.  Suppose that
$\alpha_{kl_0rs}$ is non-zero for some $l_0$ with $B_{l_0}\not=\id$.  Then
using $X= A_r\tensor B_s$ in the definition of $G_2$ and tracing out
we get $\sum_{kl}\alpha_{klrs}A_k\tensor B_l$, which is not in
$\lie{h}_a$ due to the term $B_{l_0}$.  By symmetry, this establishes
the first two identities.  The third statement follows from the
observation that any operator of the form
$(A\tensor\id)\tensor(\id\tensor B)$ is in $G_2(\lie{h}_l)$. If $A$
and $B$ are traceless, this operator is not in $\lie{h}_L$.  \qed

The above theorem provides ways of constructing $\lie{h}_A$ and
$\lie{h}_B$ but not $\lie{h}_L$.  However, one can construct
$\lie{h}_L$ as the Lie algebra generated by $\lie{h}_A$ and
$\lie{h}_B$. This depends on the bipartition only through its
emergence from having the two unilocal Lie algebras.

Another way in which one can attempt to construct $\lie{h}_X$ involves
using a group of unitary operators that extends the permutations group
$S_n$ acting on the factors.  $S_n$ by itself is insufficient, in the
sense that the Lie algebra generated by $g C g^\cstar$ for $g$ a
permutation operator and $C\in\lie{h}_{x,k}$ is just
$\bigoplus_k\lie{h}_{x,k}$. A sufficiently large extension
suffices. An example is the group $U\tensor V$, with $U$ and $V$
acting on the tensor products of the $\cH_a$ and $\cH_b$ factors,
respectively, which generates $\lie{h}_X$ from $\lie{h}_{x,1}$ by
conjugation.  The problem is whether such an extension can be chosen
naturally.  An idea that does not work but might have some independent
interest is to consider the Lie algebra $\lie{h}'_X$ generated by
$gCg^\cstar$ with $C\in\lie{h}_{x,1}$ and $g$ a unitary operator in
the group algebra generated by the permutation operators.  To see that
this does not yield the desired Lie algebras, let $s$ be the swap
operator. Then $g=(\id+is)/\sqrt{2}$ is unitary, but
$g((A\tensor\id)\tensor(\id\tensor\id))g^\cstar$ is not in
$\lie{h}_A$.

\subsection{Measures of Entanglement}
\label{sect:bi_ment}

For pure states $\ket{\psi}$ of a bipartite system the generally
accepted and information-theoretically meaningful measure of
entanglement is given by the von Neumann entropy of either one of the
reduced density matrices for $\ket{\psi}$~\cite{Donald2001a}. Thus,
the entanglement of $\ket{\psi}$ can be computed as the Shannon
entropy of the spectrum of the reduced density matrix on the first
(or, equivalently, the second) system.  For $\lie{h}_l$-states, the
underlying Hilbert space is not directly accessible. However, there
are natural complexity measures associated with the convex structure
of these states.  To define such measures, let $S$ be a Schur-concave
function of probability distributions. By definition, Schur-concave
functions are permutation invariant and concave (see for
example~\cite{Bhatia97a}, pp.~40).  That is, if $\mathbf{p}$ and
$\mathbf{q}$ are two probability distributions of the same length
where the probabilities of $\mathbf{q}$ are a permutation of those of
$\mathbf{p}$, then $S(\mathbf{p})=S(\mathbf{q})$; and if $\mathbf{p}=
r\mathbf{p}_1+(1-r)\mathbf{p}_2$ for $r\geq 0$, then
$S(\mathbf{p})\geq rS(\mathbf{p}_1)+(1-r)S(\mathbf{p}_2)$.  An example
of a Schur-concave function is the Shannon entropy.  For a pure state
$\ket{\psi}$ define $S(\ket{\psi})$ to be $S$ evaluated on the
spectrum of the reduced density matrices.  For an $\lie{h}_l$-state
$\lambda$, define
\begin{equation}
S(\lambda)=\inf\{S(\mathbf{p}) \suchthat \lambda=\sum_k p_k \lambda_k
    \; \textrm{with $\lambda_k$ $\lie{h}_l$-pure}\}.
\end{equation}
We will routinely overload the function $S$. Which definition
is intended is communicated through the argument.
So far, the argument type can be a probability distribution,
a state in $\cH_{ab}$ or an $\lie{h}_l$-state.

\begin{Theorem}
\label{thm:ent_meas1}
If the $\lie{h}_l$-state $\lambda$ is induced by a pure
state $\ket{\psi}$ on the bipartite system, then $S(\lambda)=S(\ket{\psi})$.
\end{Theorem}

\proof Using the Schmidt decomposition, we can write $\ket{\psi}=\sum_k
\sqrt{p_k}\ket{\phi_k}\tensor\ket{\varphi_k}$ with $(\ket{\phi_k})_k$ and
$(\ket{\varphi_k})_k$ orthonormal bases and
$S(\ket{\psi})=S(\mathbf{p})$.  If the $\lambda_k$ are the pure
$\lie{h}_l$-states induced by $\ket{\phi_k}\tensor\ket{\varphi_k}$,
then $\lambda=\sum_kp_k\lambda_k$. It follows that $S(\lambda)\leq
S(\ket{\psi})$.

To prove that $S(\lambda)\geq S(\ket{\psi})$, write $\lambda=\sum_k
p_k\lambda_k$, with $\lambda_k$ $\lie{h}_l$-pure and $S(\mathbf{p})$
arbitrarily close to $S(\lambda)$. To be specific, $S(\mathbf{p})\leq
S(\lambda)+\epsilon$.  By Theorem~\ref{thm:pure=ue}, the $\lambda_k$
are pure product states. Let $\lambda_k$ be induced by
$\ket{\phi_k}\tensor\ket{\varphi_k}$.  Define
$\rho=\sum_kp_k\ketbra{\phi_k}{\phi_k}\tensor\ketbra{\varphi_k}{\varphi_k}$.
Then $\trace_b(\rho)=\sum_k p_k\ketbra{\phi_k}{\phi_k}$ and is equal to
the corresponding reduced density matrix for $\ket{\psi}$. It
therefore suffices to prove that $S(\mathbf{p})$ is at least $S$
evaluated on the spectrum of $\rho_a$.  One way to see this it is to
write $\rho_a=APA^\cstar$, where $A$ consists of unit-length columns
(the $\ket{\phi_k}$) and $P$ is the diagonal matrix with the $p_k$'s
on the diagonal.  The eigenvalues of $\rho_a$ are the same as those of
$P^{1/2}A^\cstar A P^{1/2}$. This matrix has the $p_k$ on the
diagonal.  The result now follows from the fact that $\mathbf{p}$ is a
transformation of the spectrum by a doubly stochastic matrix (see, for
example,~\cite{Nielsen2001a}, page 513), doubly stochastic matrices
are convex combinations of permutation matrices (see, for
example,~\cite{Minc78a}, page 36; \cite{Nielsen2001a}, page 574), and
concavity of $S$.  \qed

Theorem~\ref{thm:ent_meas1} makes it possible to introduce
entanglement measures without reference to the underlying pair of
systems while being faithful to the known measures for such systems.
We extend the entanglement measure $S$ to mixed states by a second
minimization over convex representations as pure
states~\cite{Bennett96a}.  To do so, consider a $\lie{g}$-state
$\lambda$ induced by the density matrix $\rho$. With respect to the convex
set of $\lie{g}$-states, $\lambda$ is pure iff $\rho$ is pure. The
distinction between separability and entanglement can be seen to be
one associated with the purity of a state from the points of view of
$\lie{g}$ and $\lie{h}_l$. Thus, we define
\begin{equation}
\label{eq:convex_construction}
S(\lambda;\lie{h}_l)=
\inf\{\sum_k p_k S(\lambda_k\restrict\lie{h}_l) \suchthat
       \lambda=\sum_k p_k \lambda_k \; \textrm{with $\lambda_k$ $\lie{g}$-pure}\}
\end{equation}
Because of the isomorphism between density matrices $\rho$ and
$\lie{g}$-states, this expression defines an entanglement measure for
arbitrary bipartite density matrices. In anticipation of the
generalizations to come, we explicitly introduced the Lie algebra
$\lie{h}_l$ as a parameter.

Suppose that $S(\mathbf{p}) = 0$ iff $\mathbf{p}$ is pure, that is,
$p_k=\delta_{jk}$ for some $j$. We call
such an $S$ \emph{proper}. Then a $\lie{g}$-state $\lambda$
satisfies $S(\lambda;\lie{h}_l)=0$ iff it is a mixture of product
states, which justifies thinking of $S$ as an entanglement measure.
Several properties are desirable of an entanglement
measure~\cite{Donald2001a}. For example, the measure should be convex
and it should be non-increasing under LOCC maps. Both of these
properties are satisfied by $S$ as defined above~\cite{Vidal2000b}.

Entanglement measures can be based on asymptotic
convertibility of states with respect to a family of local maps.
For example, one can define $R(\rho,\sigma)$ as the asymptotic
supremum of $r/s$, where $r$ is the number of asymptotically good
copies of $\rho$ that can be constructed from $s$ copies of $\sigma$
given any number of additional product states and using separable
quantum maps.  For more precise definitions of this sort,
see~\cite{Bennett98a}.  If there is a reasonable choice $\sigma$ of a maximally
entangled state, then $R(\rho,\sigma)$ can be considered to be an
entropy of formation and $R(\sigma,\rho)$ an entropy of
distillation. By varying the constraints on the quantum maps
different measures are obtained.

\section{The Lie-algebraic Setting}
\label{sect:Lie}

To generalize the notions introduced in the previous section requires
not much more than removing the connection between the local Lie
algebra and the bipartite system. As a consequence we will learn that
product states are generalized coherent states.  

We fix a finite dimensional Hilbert space $\cH$ ($\cH_{ab}$ in the
bipartite setting) and consider states from the point of view of
various $\cstar$-closed, complex Lie algebras of operators acting on
$\cH$.  Ultimately, we consider families of Lie algebras
$(\lie{h}_x)_x$ acting on $\cH$ and ordered by inclusion.  But first
we consider one $\cstar$-closed Lie algebra $\lie{h}$. By default we
assume that $\id$ is a member of our operator Lie algebras.  The set
of traceless operators of $\lie{h}$ is denoted by
$\trless{\lie{h}}$. The abstract Lie algebra faithfully represented by
$\lie{h}$ is denoted by $\check{\lie{h}}$.  The assumption that
$\lie{h}$ is $\cstar$-closed implies that $\check{\lie{h}}$ is
reductive (see, for example, \cite{Dixmier96a}, Sect. 1.7). A
\emph{reductive} Lie algebra $\lie{r}$ is one that consists of the
direct product of an abelian $\lie{a}$ and a semisimple Lie algebra
$\lie{s}$ (see, for example, \cite{Dixmier96a}, Sect. 1.7, or
\cite{Humphreys72a}, p. 102).  The \emph{direct product} is in the
category of Lie algebras and homomorphisms of Lie algebras and
corresponds, after exponentiation, to the direct product of groups. In
this case it means that as vector spaces,
$\lie{r}=\lie{a}\oplus\lie{s}$, where $\lie{a}$ commutes with
$\lie{s}$. For Lie algebras, $x$ and $y$ \emph{commute} iff $[x,y]=0$.
A \emph{semisimple} Lie algebra is one which is a direct product of
simple Lie algebras, where a \emph{simple} Lie algebra is one that is
not abelian and has no proper ideals.  Reductiveness of our Lie
algebras is useful because the finite-dimensional semisimple Lie
algebras and their representations have been completely classified
(see, for example, \cite{Humphreys72a}).  If $\lie{h}$ is irreducible
as a set of operators, then the abelian part consists only of
multiples of the identity operator, and the semisimple part consists
of the traceless operators.

The two examples for $\lie{h}$ to keep in mind are $\lie{h}_l$ in the
bipartite setting and the set of generators of the spatial rotations of a
spin-$1$ particle. In the second example, the Hilbert space is three
dimensional with basis $\ket{-1}$, $\ket{0}$ and $\ket{1}$
corresponding to the three states with definite spin along $z$. The
Lie algebra $\lie{h}$ is spanned by the identity together with the
spin operators $J_z$, $J_x$ and $J_y$.  The corresponding abstract Lie
algebra is $\lie{1}\times\slc{2}$, where $\lie{1}$ is the
one-dimensional Lie algebra.  As linear spaces, this is the same as
$\lie{1}\oplus\slc{2}$, the operator $\times$ emphasizes the fact that
the construction is a direct product, so that the two Lie algebras
commute. For this example we take $\lie{g}$ to consist of all
operators.

Before proceeding, we recall the basic properties of semisimple Lie
algebras that are needed to define generalized coherent states and
relate them our characterizations of product states in the bipartite
setting. 

A Cartan subalgebra of $\trless{\lie{h}}$ is a maximal abelian
subalgebra whose elements are diagonalizable (that is,
\emph{semisimple}).  According to a fundamental result for Lie
algebras, Cartan subalgebras exist and are conjugate (hence
isomorphic) with respect to an operator in $e^{\trless{\lie{h}}}$
(\cite{Humphreys72a}, pp.~81-87; \cite{Fulton91a}, Thm. D.22, p.~492;
\cite{Varadarajan84a}, Thm. 4.1.2, p.~263).  Every diagonalizable
operator in $\lie{h}$ is contained in some Cartan subalgebra. If the
operator is hermitian, the Cartan subalgebra can be chosen to be
$\cstar$-closed.  \ignore{ 

Take any Cartan subalgebra $\lie{c}$
containing the hermitian operator $H$. Let $\cB$ be a Borel subalgebra
for $\lie{c}$. Then $H\in\cB\cap\cB^\cstar$. The point is that
$\lie{c}'=\cB\cap\cB^\cstar$ is a Cartan subalgebra that is
$\cstar$-closed. This is because $\lie{c}'$ is solvable (because it is
a subalgebra of a solvable Lie algebra) and reductive (because it is
$\cstar$-closed). This implies that it is abelian.  A dimension
argument shows that it has to have the same dimension as $\lie{c}$.  

}%
Let $\lie{c}$ be a Cartan subalgebra of $\trless{\lie{h}}$, then $\cH$
can be decomposed into the joint eigenspaces for $\lie{c}$,
$\cH=\bigoplus_\alpha\cH_\alpha$, where the $\alpha$ are distinct
linear functionals on $\lie{c}$ such that for
$\ket{\psi}\in\cH_\alpha$ and $A\in \lie{c}$,
$A\ket{\psi}=\alpha(A)\ket{\psi}$.  (\cite{Humphreys72a}, p.~107;
\cite{Fulton91a}, p.~199 eq. (14.4)).  The $\cH_\alpha$ are called the
\emph{weight spaces} for $\lie{c}$ and the $\alpha$ are called the
\emph{weights}. In general, a weight for a Cartan subalgebra is a
linear functional for which there exists a finite dimensional
representation with a non-empty corresponding weight space.  The
abstract Lie algebra $\check{\trless{\lie{h}}}$ can be represented on
itself by the Lie bracket. This is called the \emph{adjoint}
representation of $\check{\trless{\lie{h}}}$. The weights for this
representation are called \emph{roots}.  It turns out that the
geometrical properties of the roots determine the Lie algebra. The
roots are in effect also linear functionals on $\lie{c}$. There are
special sets of roots called \emph{simple root systems} (or
\emph{bases}) that span the linear functionals on $\lie{c}$ and have
the property that every root is either a positive, or a negative
integral combination of simple roots. The former are called
\emph{positive} roots. The definition depends on the choice of simple
roots, but not in a crucial way, because all simple root systems are
isomorphic via a special kind of isomorphism (a member of the
so-called Weyl group, \cite{Humphreys72a}, p.~51; \cite{Fulton91a},
Prop.~D.29, p.~494).  The weights can be partially ordered by defining
$\alpha\leq\alpha'$ if $\alpha'-\alpha$ is a positive integral sum of
simple roots.  With this ordering, in an irreducible representation,
there is a unique minimum weight, whose weight space is
one-dimensional (\cite{Humphreys72a}, pp.~108--109; \cite{Fulton91a},
Prop.~14.13, pp.~202--203).  The minimum weight state depends on the
choice of Cartan subalgebra and simple roots.  However,
$e^{i\real(\trless{\lie{h}})}$ acts transitively on the set of minimum
weight vectors. Furthermore, every minimum weight vector can be
obtained by means of a $\cstar$-closed Cartan subalgebra of $\lie{h}$.
The minimum weight space has the property that it is annihilated by
operators in $\trless{\lie{h}}$ which are in root spaces associated
with negative roots. In fact, this is another characterization of the
minimum weight space (see the definition and theorem in
\cite{Humphreys72a}, p.~108).  Usually, treatments of semisimple Lie
algebras focus on the maximum weights of a representation. Here we
choose to use the equivalent minimum weights because of the
relationship to ground states of Hamiltonians.  The basic properties
of Cartan subalgebras and the notions of roots and weights extend from
semisimple to reductive Lie algebras by adjoining the abelian part.
\ignore{

We also need to refer to \emph{Borel subalgebras}, which are maximal
solvable subalgebras of a Lie algebra (\cite{Humphreys72a},
p.~83). Like the Cartan subalgebras, any two Borel subalgebras are
conjugate.  In this case, the Lie algebra need not be semisimple or
reductive. Given a Cartan subalgebra $\lie{c}$ with an ordering of the
roots determined by a simple root system, a Borel subalgebra $\lie{b}$
consists of the linear span of the negative roots together with
$\lie{c}$.  If $\lie{c}$ is $\cstar$-closed, then
$\lie{b}+\lie{b}^\cstar=\trless{\lie{h}}$ and
$\lie{b}\cap\lie{b}^\cstar=\lie{c}$. Minimum weight states can be
characterized as eigenvectors of Borel subalgebras.

}%

A family of generalized coherent states consists of an orbit of a
dynamical group acting on a state space~\cite{Perelomov85a,Zhang90a}.
According to this definition, every state is in a family of
generalized coherent states. As a result, an important part of the
theory of generalized coherent states is to choose those orbits that
best generalize the properties of the coherent states familiar in
optics. In our case, the dynamical groups are Lie groups generated by
semisimple Lie algebras.  If the goal is to choose states that are in
a sense the most classical, then there are strong arguments for
choosing the minimum weight states of a representation of the Lie
group. Theorem~\ref{thm:stchar} below provides some of these
arguments. We therefore use the term \emph{generalized coherent}
state, or simply \emph{coherent} state, to refer specifically to
minimum weight states of a Lie algebra.  Because we only consider
finite dimensional representations, our treatment does not directly
apply to the conventional coherent states of optics, for example. In
this case, the relevant Lie algebra is the Heisenberg algebra, which
is not semisimple (or reductive). The standard $\cstar$-closed
representation is therefore necessarily infinite.  The theory of
coherent states suggests that extensions to such Lie algebras and
representations are possible~\cite{Perelomov85a}.

\subsection{Purity, Coherence and Entanglement}

For a $\cstar$-closed Lie algebra of operators $\lie{h}$ on $\cH$,
define $\lie{h}$-states as before as linear functionals on $\lie{h}$
induced by a state's density matrix $\rho$ according to
$\lambda(A)=\trace(\rho A)$. Observe again that the set $\lie{h}^{+}$
of $\lie{h}$-states is convex closed. \emph{Pure} $\lie{h}$-states are
extreme points of $\lie{h}^{+}$. Suppose that the $\lie{h}$-state
$\lambda$ is induced by the density matrix $\rho$. We can project
$\rho$ onto $\lie{h}$ with respect to the trace inner product.  Denote
the projection map onto $\lie{h}$ by $\cP_{\lie{h}}$. Because
$\lie{h}$ is $\cstar$-closed, the projection $\cP_{\lie{h}}(\rho)$ is
a hermitian operator in $\lie{h}$. Furthermore, $\lambda$ is also
induced by $\cP_{\lie{h}}(\rho)$, that is,
$\lambda(A)=\trace(P_{\lie{h}}(\rho) A) = \trace(\rho A)$ for
$A\in\lie{h}$. Note that in general, $\cP_{\lie{h}}(\rho)$ is not
positive.  For example, let $\rho$ be the density matrix for $\ket{1}$
in the spin-$1$ example. Another important observation is that
$\cP_{\lie{h}}(\rho)$ depends only on $\lambda$.  That is, if $\rho$
and $\rho'$ both induce $\lambda$, then
$\cP_{\lie{h}}(\rho)=\cP_{\lie{h}}(\rho')$.

We now assume that $\lie{h}$ acts irreducibly on $\cH$. If it does not
act irreducibly, decompose $\cH$ into irreducible invariant subspaces
for $\lie{h}$ and consider each of these subspaces separately.  Define
the \emph{$\lie{h}$-purity} of $\lambda$ as
$\trace(\cP_{\lie{h}}(\rho)^2)$, where $\lambda$ is induced by the
density matrix $\rho$.  This is of course the length of
$\cP_{\lie{h}}(\rho)$ according to the trace-inner-product norm.  The
$\lie{h}$-purity is bounded above by the conventional purity
$\trace(\rho^2)$, which is the $\lie{g}$-purity with $\lie{g}$ the
algebra of all operators on $\cH$. This generalization of purity is
useful because according to Theorem~\ref{thm:stchar} below, the pure
$\lie{h}$-states are exactly the states with maximum $\lie{h}$-purity.

The goal of the remainder of this subsection is to give a number of
useful characterizations of pure $\lie{h}$-states.  In particular, we
show that they are exactly the coherent states for $\lie{h}$.  We
first state the characterization theorem and then discuss the
equivalent characterizations before proving the theorem.

\newcounter{stcharcnt}
\begin{Theorem} 
\label{thm:stchar}
The following are equivalent for a density matrix $\rho$
inducing the $\lie{h}$-state $\lambda$:
\begin{itemize}

  \refstepcounter{stcharcnt}\label{st1}
  \item[\emph{(\textbf{\thestcharcnt})}] $\lambda$ is a pure $\lie{h}$-state.

  \refstepcounter{stcharcnt}\label{st2}
  \item[\emph{(\textbf{\thestcharcnt})}] $\rho=\ketbra{\psi}{\psi}$ with 
  $\ket{\psi}$ the unique ground state of 
  some $H$ in $\real(\lie{h})$.

  \refstepcounter{stcharcnt}\label{st3}
  \item[\emph{(\textbf{\thestcharcnt})}] $\rho=\ketbra{\psi}{\psi}$ with
  $\ket{\psi}$ a minimum-weight vector (for some simple root system
  of some Cartan subalgebra) of $\trless{\lie{h}}$.

\ignore{
  \refstepcounter{stcharcnt}\label{st4}
  \item[\emph{(\textbf{\thestcharcnt})}] $\rho=\ketbra{\psi}{\psi}$ such that
  $R=\{r\in\trless{\lie{h}} \suchthat r\ket{\psi}=\lambda_r\ket{\psi}\}$ 
  satisfies $R+R^\cstar=\trless{\lie{h}}$.
}%

  \refstepcounter{stcharcnt}\label{st5}
  \item[\emph{(\textbf{\thestcharcnt})}] $\lambda$ has maximum $\lie{h}$-purity.

  \refstepcounter{stcharcnt}\label{st6}
  \item[\emph{(\textbf{\thestcharcnt})}] $\rho$ is a one-dimensional projector in
   $\overline{e^{\lie{h}}}$.
\end{itemize}
\end{Theorem}

This theorem is a synthesis of various largely known results in the
representation theory of semisimple Lie algebras and coherent states.
Statements (\textbf{\ref{st1}}) and (\textbf{\ref{st2}}) are motivated by
Theorems~\ref{thm:pure=ue} and~\ref{thm:prod=ue}, respectively.
(\textbf{\ref{st2}}) also provides an interpretation of many meanfield ground
states as coherent states. This is because meanfield Hamiltonians are
often expressed as operators in a small Lie algebra, in particular,
operators quadratic in the creation and annihilation operators.

Statement (\textbf{\ref{st3}}) is one of the definitions of
generalized coherent states. For other characterizations of
generalized coherent states, see~\cite{Perelomov85a,Zhang90a}.

\ignore{
Statements (\textbf{\ref{st3}}) and (\textbf{\ref{st4}}) are two equivalent
definitions of generalized coherent
states~\cite{Perelomov85a,Zhang90a}. According to (\textbf{\ref{st4}}),
coherent states are those with a lot of symmetries, that is, the
states $\ket{\psi}$ for which $G_{\ket{\psi}}=\{U\in
e^{\trless{\lie{h}}}\suchthat U\ket{\psi} \propto \ket{\psi}\}$ is
large.  $G_{\ket{\psi}}$ is called the \emph{stabilizing} or
\emph{isotropy} subgroup for $\ket{\psi}$.  It can be shown that
(\textbf{\ref{st4}}) implies that the stabilizing subgroup is maximal.
}%

Statement (\textbf{\ref{st5}}) is a version of the minimum variance
principle for coherent states~\cite{Delbourgo77a,Delbourgo77b}. The
variance of a state $\ket{\psi}$ with respect to
$\real(\trless{\lie{h}})$ is computed as the expectation of an
``invariant uncertainty operator''. For a state $\ket{\psi}$, this
expectation is given by
\begin{equation}
\label{dispersionop}
\sum_i x^i x_i - \sum_i \bra{\psi} x^i \ketbra{\psi}{\psi} x_i \ket{\psi},
\end{equation}
where $(x_i)_i$ is a basis of $\real(\trless{\lie{h}})$, and $(x^i)_i$
is the dual basis with respect to the trace inner product. This is a
linear function of the $\lie{h}$-purity because the second sum is the
negative of the purity up to a constant due to our inclusion of the
identity operator.

Statement (\textbf{\ref{st6}}) is motivated by the results concerning the
classical simulatability of fermionic linear
optics~\cite{Terhal2001a,Knill2001b}. Simulatability depends crucially
on the fact that the initial state preparations and the measurements
outcomes can be expressed in terms of projectors in
$\overline{e^{\lie{h}}}$.

\proofof{\ref{thm:stchar}} 

(\textbf{\ref{st2}})~$\Rightarrow$~(\textbf{\ref{st3}}): Let $\lie{c}$ be a
$\cstar$-closed Cartan subalgebra of $\trless{\lie{h}}$ containing
$H$. We can perturb $H$ slightly without affecting the ground space by
adding a generic element of $\lie{c}$ to make sure that $H$ is
generic, that is, so that the commutant of $H$ is $\lie{c}$.  The
\emph{commutant} of $H$ is the set of elements of $\trless{\lie{h}}$
that commute with $H$.  It therefore suffices to show that ground
states of generic elements of $\lie{c}$ are minimum weight for an
ordering of the roots. Note that for no non-zero root $\alpha$ is
$\alpha(H)=0$, because otherwise $H$ is not generic. Thus we can call
a root positive if $\alpha(H)>0$, and there is some simple root system
for which this coincides with the definition of positive roots.  A
ground state is annihilated by the root spaces of
$\trless{\lie{h}}$ that correspond to the negative roots. This implies
that it is a minimum weight state.

(\textbf{\ref{st3}})~$\Rightarrow$~(\textbf{\ref{st2}}): Every minimum-weight vector
$\ket{\psi}$ has minimum weight for some $\cstar$-closed Cartan
subalgebra $\lie{c}$ with root basis $\alpha_1,\ldots,\alpha_d$. There
is a hermitian member $H$ of $\lie{c}$ for which $\alpha_k(H)>0$ for
each $k$.  $\ket{\psi}$ is the ground state of $H$.

\ignore{
(\textbf{\ref{st3}})~$\Rightarrow$~(\textbf{\ref{st4}}): $R$ contains the Cartan
subalgebra together with the root spaces for negative roots.  We can
assume, without loss of generality, that the Cartan algebra is
hermitian, so $R^\cstar$ consists of the Cartan algebra together with
the positive roots, and (\textbf{\ref{st4}}) follows.
}%

(\textbf{\ref{st2}})~$\Rightarrow$~(\textbf{\ref{st6}}): Let $\lambda$ be the eigenvalue
of $\ket{\psi}$ for $H$.  Then the desired projector is
$\lim_{t\rightarrow \infty} e^{(-H+\lambda)t}$.

((\textbf{\ref{st6}})~\&~((\textbf{\ref{st2}})~$\Rightarrow$~(\textbf{\ref{st3}})))~$\Rightarrow$~(\textbf{\ref{st3}}):
Let $\ketbra{\psi}{\psi}=\lim_k e^{-H_k}$, with $H_k\in\lie{h}$.  Then
$\ketbra{\psi}{\psi}=\lim_k e^{-H_k}e^{-\scstar{H_k}}$.  The operators
in the limit are now hermitian, which implies that they can be written
in the form $e^{-h_k}$, with $h_k$ hermitian in $\lie{h}$.  
For sufficiently large $k$, the
minimum eigenvalue of $h_k$ must be unique. This eigenvalue
must go to zero and the eigenvalue gap $\delta_k$ of $h_k$
goes to infinity.  Thus for sufficiently
large $k$, the ground state $\ket{\psi_k}$ of $h_k$ is projectively
well-defined. Because of ((\textbf{\ref{st2}})~$\Rightarrow$~(\textbf{\ref{st3}}))),
$\ket{\psi_k}$ is a minimum weight state. Minimum weight states
form an orbit of $e^{\real(\lie{h})}$, a compact set.
Thus there is a cluster point $\ket{\psi_0}$ of the $\ket{\psi_k}$.
It must be the case that $\ket{\psi_0}\propto\ket{\psi}$.
Hence $\ket{\psi}$ is minimum weight.

(\textbf{\ref{st5}})~$\Rightarrow$~(\textbf{\ref{st1}}): By convexity of purity.

(\textbf{\ref{st1}})~$\Rightarrow$~(\textbf{\ref{st3}}): Let $\lie{c}$ be the
$\cstar$-closed Cartan subalgebra containing the projection of $\rho$
into $\lie{h}$.  We call this a \emph{supporting} Cartan subalgebra of
$\rho$. Let $\cH_\alpha$ be the weight spaces with respect to this
Cartan subalgebra.  Then $\lambda$ is zero on the non-zero root spaces
with respect to $\lie{c}$.  Since $\rho$ is a mixture of normalized
superpositions of weight vectors $\ket{v_\alpha}\in\cH_{\alpha}$, it
follows that $\lambda\restrict\lie{c}$ is a convex combination of
weights. But the weights are all in the convex closure of the set of
minimum weights with respect to different orderings of the roots.
Extremality therefore requires that $\lambda\restrict \lie{c}$ is
given by a minimum weight. Let $\ket{\psi}$ be the corresponding
minimum weight state.  By choice of $\lie{c}$, $\lambda$ is also
induced by $\ketbra{\psi}{\psi}$.  The density matrix $\rho$ cannot
have a contribution to the mixture with different weight spaces, as
otherwise, $\lambda\restrict\lie{c}$ is in the strict interior of the
convex closure of the set of minimum weights. That
$\rho=\ketbra{\psi}{\psi}$ now follows from the fact that due to
irreducibility of $\lie{h}$, the minimum weight spaces are
one-dimensional.

Note that supporting Cartan subalgebras' weight spaces generalize
the Schmidt basis used to diagonalize reduced density matrices
in the bipartite setting. See Theorem~\ref{thm:supp_cartan}.

((\textbf{\ref{st1}})~\&~(\textbf{\ref{st3}}))~$\Rightarrow$~(\textbf{\ref{st5}}): Because all
minimum weight states are in the same orbit of $e^{i\real(\lie{h})}$,
every minimum weight state has the same purity. By extremality and
convexity of purity, minimum weight states have maximum purity.

\ignore{
(\textbf{\ref{st4}})~$\Rightarrow$~(\textbf{\ref{st3}}): 
\mComment{To be completed.}
\ignore{
Let $\lie{c}$ be the supporting
Cartan subalgebra of $\rho$.  Let $\alpha$ be the weights of $\lie{c}$
and $\cH_\alpha$ the corresponding weight spaces. For each root
$\beta$ of $\lie{c}$, let $x_\beta$ be a non-zero element of the root
space for $\beta$. The $x_\beta \cH_{\alpha}\subseteq
\cH_{\alpha+\beta}$ Write $\ket{\psi} =
\sum_{\alpha}a_\alpha\ket{\psi_\alpha}$ with $\ket{\psi_\alpha}\in
H_\alpha$.  By the definition of supporting Cartan subalgebras,
$\trace(x_\beta\rho)=0$ for each root $\beta$.  Let $s\in\lie{h}$ such
that $s\ket{\psi}=\gamma\ket{\psi}$.  Write $s=c+\sum_{\beta} b_\beta
x_\beta$ with $c\in\lie{c}$. 
}%
}%

\qed

\subsection{Local Quantum Maps}
\label{sect:li_lqo}

We can use Theorem~\ref{thm:clieh=sep} to generalize separable
maps. Thus we define \emph{$\lie{h}$-separable} quantum maps to be
those with an explicit form $(A_k)_k$ with
$A_k\in\overline{e^{\lie{h}}}$.  To generalize LOCC maps, one can
always return to the multipartite setting by using the fact that by
semisimplicity, $\trless{\lie{h}}$ can be uniquely represented as a
product of simple Lie algebras $\trless{\lie{h}}=\times_k\lie{h}_k$
(see, for example, \cite{Humphreys72a}, p. 23).
The state space then factors as $\tensor_k\cH_k$, with $\lie{h}_k$
acting on $\cH_k$ only. We define $\lie{h}$-LOCC maps by
conditional composition of explicit $(\lie{h}_k+\cmplx\id)$-separable
quantum maps. This definition is more general than the usual notion of
LOCC maps for multipartite systems because $\check\lie{h}_k$ can be
different from $\slc{n}$ or its representation $\lie{h}_k$ need not be
the first fundamental representation.

In the bipartite setting, we discussed two other ways in which LOCC
maps can be characterized. One way used liftability to well defined
maps of $\lie{h}$-states. The other used restrictions on the operators
based on their eigenspaces. We consider how these ideas can lead to
other interesting families of quantum maps.

A subfamily of the explicit $\lie{h}$-separable quantum maps is
obtained by requiring that each operator lifts to $\lie{h}$.  Such
quantum maps are called \emph{explicit $\lie{h}$-liftable} quantum
maps.  (Recall our convention for using the word ``explicit''.)  In
the bipartite setting, Theorem~\ref{thm:liftone_diu} implies that all
such quantum maps are mixtures of unitaries, a small subfamily of the
LOCC maps.  The conclusion of Theorem~\ref{thm:liftone_diu} does not
hold in general. For example, trivially, if $\lie{h}$ consists of all
operators on $\cH$, then all quantum maps are in this family. One nice
property of the family of explicit $\lie{h}$-liftable quantum maps is
that there is a straightforward proof of monotonicity for a large
family of entanglement measures, see Theorem~\ref{thm:monotonicity}.

A family of quantum maps that includes the explicit $\lie{h}$-liftable
ones consists of the $\lie{h}$-separable quantum maps that are
liftable to $\lie{h}$.  In the bipartite setting, this family may be
larger than the family of LOCC maps, see
Problem~\ref{prob:sep_lift}. In the general setting, we pose the
following problem:

\begin{Problem}
Is the family of quantum maps obtained by conditional
composition of explicit $\lie{h}$-separable quantum maps that are
liftable to $\lie{h}$ strictly smaller than the family
of $\lie{h}$-separable quantum maps?
\end{Problem}

Based on Theorem~\ref{thm:nsep_locc} and its corollary, one might want
to consider the family of maps consisting of binary
$\lie{h}$-separable quantum maps. Unfortunately, this family can be
trivial in the sense that in many cases it consist of mixed unitary
quantum maps only.  For example, consider the spin-$1$ Lie algebra and
suppose that $(A,B)$ is an explicit separable
quantum map. We have $A,B\in e^{\lie{h}}$ and $A^\cstar A+B^\cstar
B=\id$. The operators $A^\cstar A$ and $B^\cstar B$ are in $e^{\lie{h}}$ and
can be written in the form $e^{H_A}$ and $e^{H_B}$ with $H_A$ and
$H_B$ in $\real(\lie{h})$.  Thus $H_A=\alpha\id+\vec x \cdot \vec
J$. With a suitable rotation, we can assume that $H_A=\alpha+\beta
J_z$. This ensures that $e^{H_A}$ is diagonal in the basis
$\ket{-1},\ket{0},\ket{1}$ and has diagonal entries
$e^{\alpha-\beta},e^{\alpha},e^{\alpha+\beta}$.  It follows that
$e^{H_B}$ is diagonal also, and hence of the same form with $\alpha'$
and $\beta'$. Their sum is $\id$, and it can be checked that the
solutions satisfy $\beta=\beta'=0$. Hence $A$ and $B$ are proportional
to unitaries.

\ignore{
(* Find positive solution for a,b,x,y in: *)
eq1=(a+b==1);
eq2=(a*x+b*y==1);
eq3=(a/x+b/y==1);
sol=Solve[eq1&&eq2&&eq3,{a,b,x,y}];
(* Only ``trivial'' solutions. *)
sola=Solve[eq1,{a}];
eq2a=eq2/.sola[[1]]; eq3a=eq3/.sola[[1]];
solb=Solve[eq2a,{x}];
(* b \not= 1, but then a==0. *)
eq3b=eq3a/.solb[[1]];
(* eq3b:
b/y + (1 - b)^2/(1 - b*y) == 1
* The two solutions to this both have y==1.
* So same goes for x.
*)
}%

One idea for avoiding the possible triviality of binary
$\lie{h}$-separable quantum maps is to use $k$-ary quantum maps. That
is, consider extremal $k$-ary $\lie{h}$-separable quantum maps. A
quantum map is \emph{extremal} if its action on density matrices is
not a convex combination of other quantum maps.  Because mixed unitary
quantum maps are not extremal unless they are unitary, the spin-$1$
example shows that there may be no such extremal quantum maps for
$k=2$. Let $k_{\textrm{\tiny min}}$ be the minimum $k>1$ for which
such quantum maps exist.  Let the family of \emph{minimally generated}
separable quantum maps consist of explicit quantum maps obtained by
conditional composition of unary or extremal $k_{\textrm{\tiny
min}}$-ary $\lie{h}$-separable quantum maps.  Because of
Corollary~\ref{cor:2sep_locc}, this family is the family of LOCC maps
in the bipartite setting.

\begin{Problem}
What is the relationship between the family of minimally generated
$\lie{h}$-separable quantum maps, $\lie{h}$-\emph{LOCC} and 
and $\lie{h}$-separable quantum maps?
\end{Problem}

Another family of quantum maps that might be interesting is
obtained by adding the liftability condition to the 
generators of the family in the above problem.

We now move on to considering families of $\lie{h}$-separable quantum
maps that are characterized by generators with large ground
spaces. Based on Theorem~\ref{thm:maxeig=unil}, we can define a
\emph{maximally $\lie{h}$-unilocal} operator to be an operator in
$\real(\lie{h})$ whose ground space is maximal. These operators have a
Lie algebraic characterization.

\begin{Theorem}
\label{thm:maxeig_char}
Maximally $\trless{\lie{h}}$-unilocal operators are the ones that are
proportional to a an operator of the dual
basis to a simple root system  of a $\cstar$-closed Cartan
subalgebra of $\trless{\lie{h}}$.
\end{Theorem}

The dual basis of a simple root system corresponds to the fundamental
weights via the isomorphism induced by the Killing form. The Killing
form is the symmetric bilinear form associated with the trace in the
adjoint representation.  The $k$'th fundamental weight $\lambda_k$ for
a simple root system consisting of the roots $\alpha_l$ has the
property that if $h_l=[x_l,y_l]$ with $x_l$ and $y_l$ members of the
root space for $\alpha_l$ and for $-\alpha_l$, respectively, then
$\lambda_k(h)=0$ except for $l=k$. It also has minimum length among
weights satisfying this property.  Fundamental weights are important
because all the representations of a Lie algebra can be built from
ones whose minimum weight is fundamental.

\proofof{\ref{thm:maxeig_char}} Let $H\in\real(\lie{h})$ and choose a
$\cstar$-closed Cartan subalgebra $\lie{c}$ containing $H$ and an
ordering of the roots such that for positive roots $\alpha$,
$\alpha(H)\geq 0$.  Let $(\alpha_k)_k$ be the simple root system for
this ordering.  Let $\cH_0$ be the ground space of $H$. Then $\cH_0$
is a union of weight spaces of $\lie{c}$. By definition of the ground
space, if $X$ is in the root space for a negative root, then
$X\cH_0\subseteq\cH_0$.  In particular, $\cH_0$ contains the weight
space for the minimum weight $\lambda_0$ of the chosen ordering of the
roots.  Furthermore, $\cH_0$ consists exactly of the weights $\lambda$
such that $\lambda-\lambda_0$ is a positive integral combination of
positive roots $\alpha$ with $\alpha(\cH_0)=0$.  Thus $\cH_0$ is
non-trivially maximal iff $\alpha_k(\cH_0)=0$ for all but one
$k=k_0$. Given $k_0$, the set of operators with this property is
necessarily one-dimensional and contains one that contributes to the
dual basis of the simple root system. This follows from the fact that
the simple roots are a basis of the dual space of $\lie{c}$.  \qed

The maximally $\lie{h}$-unilocal operators fall into different classes
depending on the associated fundamental weight.  However, it is likely
that if $\trless{\lie{h}}$ is simple, then the linear span of the
$e^{\lie{h}}$ conjugates of a given maximally
$\trless{\lie{h}}$-unilocal operator is all of $\trless{\lie{h}}$. We
do not know whether this holds in general, but it is certainly the
case for $\lie{h}_a$ and $\lie{h}_b$ and $\lie{g}$.  This implies that
if we define m-compatibility as in the bipartite setting and close
under conditional composition, we might get all $\lie{h}$-LOCC maps. So
define a \emph{$\lie{h}$-compatible} family of operators as a family
consisting of the $e^{\lie{h}}$ conjugates of a maximally
$\lie{h}$-unilocal operator.

\begin{Problem}
Does conditional composition of explicit separable quantum
maps with operators from an $\lie{h}$-compatible family
generate the family of $\lie{h}$-\emph{LOCC} maps?
\end{Problem}

For now, the properties of the various families of quantum maps are
largely unknown and offer a fruitful area of further investigation.

\subsection{Communication Complexity}

Communication complexity can be defined exactly as in the bipartite
setting for any of the families of explicit quantum maps
defined by conditional composition in the previous section.

\subsection{Resource Scaling}

The goal is to determine what might be reasonable choices of
``scaled'' Lie algebras $\lie{h}^{\circledast n}$ acting on
$\cH^{\tensor n}$ extending the action of $\lie{h}$ on each factor so
as to be consistent with the corresponding picture for bipartite
systems. It makes sense to require that $\lie{h}^{\circledast n}$ be
contained in $G_n(\lie{h})$, the set of operators $X$ with the
property that if $Y$ is an operator acting as the identity on the
$k$'th factor of $\cH^{\tensor n}$, then the partial trace of $XY$
onto the $k$'th factor is in $\lie{h}$ acting on this factor. In the
bipartite case, it was possible to obtain the desired
$\lie{h}^{\circledast n}$ by appealing to the two unilocal Lie
algebras contained in $\lie{h}$. We can similarly use any generating
Lie subalgebras.  That is, let $\lie{h}$ be generated by Lie
subalgebras $\lie{h}_k$.  With respect to these Lie subalgebras, we can
define $\lie{h}^{\circledast n}$ as the Lie algebra generated by
$G_n(\lie{h}_k)$. In this case, it makes sense to define
${\lie{h}_k}^{\circledast n}=G_n(\lie{h}_k)$.  At this point, we do
not know to what extent this scheme is useful in analyzing the
asymptotic relationships between states from the point of view of
$\lie{h}$.  As a potentially interesting alternative, the scheme based
on extensions of the permutation group discussed in the last paragraph
of of Section~\ref{sect:bi-resources} can of course be applied to any
Lie algebra of operators.

\subsection{Measures of Relative Entanglement}

From the point of view of $\lie{h}$, incoherent pure states of $\cH$
look like a mixture of coherent states. This is because the
$\lie{h}$-state induced by an incoherent state is a proper convex
combinations of pure $\lie{h}$-states.  However, incoherent pure
states can exhibit generalized entanglement provided that it is
possible to refer to operators outside of $\lie{h}$.  We therefore
need access to observables in a larger Lie algebra.  Let
$\lie{g}\supset\lie{h}$ be a Lie algebra of operators on
$\cH$. Theorem~\ref{thm:stchar} applies to $\lie{g}$ as well, and in
general, not all pure $\lie{g}$-states are pure when restricted to
$\lie{h}$. Note that a $\lie{g}$-state that restricts to a pure
$\lie{h}$-state state is necessarily pure. So it makes sense to call a
pure $\lie{g}$-state \emph{$\lie{h}$-coherent} if it restricts to a
pure $\lie{h}$-state.

The goal of this section is to find ways to quantify the
\emph{relative entanglement} of $\lie{g}$-states with respect to
$\lie{h}$. The idea is that $\lie{h}$-coherent $\lie{g}$-states are
not entangled, while any other pure $\lie{g}$-state is definitely
entangled, but the extent of entanglement depends in some way on how
far the state is from being pure when restricted to $\lie{h}$. Once
the entanglement of pure $\lie{g}$-states has been quantified, this
can be extended to arbitrary $\lie{g}$-states.

Let $S$ be a Schur-concave function of probability distributions.
Then we can define $S(\lambda)$ for $\lie{h}$-states $\lambda$ and
$S(\lambda';\lie{h})$ for $\lie{g}$-states $\lambda'$ as we did in
Section~\ref{sect:bi_ment}.  In the bipartite setting, $S(\lambda)$ is
concave as a function of $\lie{h}$-states $\lambda$.

\begin{Problem}
For which $\lie{h}$ is $S$ a concave function of $\lie{h}$-states?
\end{Problem}

That $S(\lambda;\lie{h})$ is a convex function of $\lie{g}$-states
$\lambda$ will be shown in the more general setting of convex cones,
where we will also discuss the issue of monotonicity of $S$ under the
various notions of generalized local quantum maps.

Another measure that can be used for quantifying generalized
entanglement is based on purity. Let $p(\lambda')$ denote
the $\lie{h}$-purity of an $\lie{h}$-state $\lambda'$.
We can define, for a $\lie{g}$-state $\lambda$,
\begin{equation}
p(\lambda;\lie{h})=
\sup\{\sum_k p_k p(\lambda_k\restrict\lie{h}) \suchthat
       \lambda=\sum_k p_k \lambda_k \; \textrm{with $\lambda_k$ pure for $\lie{g}$}\}
\end{equation}
Then $p(\lambda;\lie{h})$ achieves its maximum exactly at the states
that are mixtures of $\lie{h}$-coherent states, and
$p(\lambda;\lie{h})$ is convex in $\lambda$.  Mixtures
of $\lie{h}$-coherent states are generalized separable states.

Observe that for bipartite pure states, the purity is a
linear function of the Renyi entropy given by $-\sum_k p_k^2$ where
the $p_k$ are the eigenvalues of the reduced density matrices. In this
case, the Renyi entropy can be derived from the Schur-concave function
$S((p_k)_k)=-\sum_k p_k^2$.

It is possible to define resource-based measures of relative
entanglement as discussed at the end of Section~\ref{sect:bi_ment},
with the caution that asymptotic versions of such measures depend on
whether a useful notion of scaling for resources can been found.

One advantage of relativizing measures of entanglement by using pairs
$\lie{h}\subseteq\lie{g}$, is that one can better investigate
properties of states on systems with a hierarchy of meaningful choices
for Lie algebras.  Multipartite systems are examples where this
situation arises.  For every subset $s$ of the subsystems, there is
the algebra $\lie{h}_s$ of operators acting only on the subsystems in
$s$, and the $\lie{h}_s$ can be summed over a partition of the
subsystems to obtain generalizations of $\lie{h}_l$.  These Lie
algebras are ordered by inclusion. Given a state, one can, for every
pair $\lie{k}\subseteq\lie{l}$, determine the state's generalized
entanglement and use these quantities to characterize different types
of states and localize the extent to which they are entangled.
Other examples with multiple, physically motivated Lie algebras 
are discussed in Section~\ref{sect:further}.

\subsection{Other Measures}
\label{sect:lie_om}

We mention two other types of relative entanglement measures for
states that may generalize the bipartite setting. One is based on the
amplitudes in a representation of a state as a superposition of
coherent states, the other uses supporting Cartan subalgebras as a
generalization of the Schmidt basis. Since both of them can be
extended to mixed $\lie{g}$-states using the construction repeatedly
used above (see Equation~\ref{eq:convex_construction}), we discuss
them only for pure $\lie{g}$-states. Since these are induced by pure
states of $\cH$ and the relativization comes in through the extension,
we define the measures for all pure states $\ket{\psi}\in\cH$.

Let $S$ be a Schur-concave function and $\ket{\psi}$ a state that
induces a pure $\lie{g}$-state. We can define an entanglement
measure by minimizing the $S$-complexity of $\ket{\psi}$'s
renormalized square amplitudes in writing $\ket{\psi}$ as a
superposition of coherent states.  Formally:
\begin{equation}
S_a(\ket{\psi})
 = \inf\{S(\mathbf{p}) \suchthat
       p_k = |\alpha_k|^2/\sum_k |\alpha_k|^2\;\textrm{where}\;
       \ket{\psi}= \sum_k \alpha_k\ket{\psi_k}\;\textrm{with}\;
       \textrm{$\lie{h}$-coherent $\ket{\psi_k}$}.
       \}
\end{equation}
Note that by irreducibility of $\lie{h}$, every state is in the span
of the coherent states for $\lie{h}$. 

\begin{Problem}
Is $S_a(\ket{\psi}) = S(\ket{\psi})$ in the bipartite setting? 
\end{Problem}

$S(\ket{\psi})$ is defined for the bipartite setting before
Theorem~\ref{thm:ent_meas1}.

A limiting case of this definition is the $\lie{h}$-rank of
$\ket{\psi}$ defined as the minimum number of states needed to
represent $\ket{\psi}$ as a superposition of coherent states. The
$\lie{h}$-rank is obtained as the limit of the Schur-concave functions
$S_r:\mathbf{p}\rightarrow \sum_k p_k^{1/r}$ as $r\rightarrow \infty$.
A special case of the $\lie{h}$-rank has a long history in quantum
chemistry (see, for example~\cite{Mcweeny92a}, p.~69) and has been
proposed in the context of entanglement for fermions
in~\cite{Schliemann2001a}, and for bosons
in~\cite{Paskauskas2001a,Eckert2002a}.

\begin{Problem}
What is the relationship between the amplitude-based
($S_a(\ket{\psi})$) and the convexity-based ($S(\lambda)$) measures of
entanglement for pure $\lie{g}$-states?
\end{Problem}

$S_a$ satisfies that for proper $S$, $S_a(\ket{\psi})=0$
iff $\ket{\psi}$ is coherent for $\lie{h}$. 
The measure $S_C(\ket{\psi})$ based
on supporting Cartan subalgebras does not satisfy this.
To define $S_C(\ket{\psi})$, let $\lie{c}$ be a 
supporting Cartan subalgebra of $\lie{h}$ for $\ketbra{\psi}{\psi}$.
Let $P_\alpha$ be the projectors onto the weight spaces
of $\lie{c}$. We can define 
\begin{equation}
\label{eq:S_C}
S_C(\ket{\psi}) = \inf S((|P_\alpha\ket{\psi}|^2)_\alpha),
\end{equation}
where the minimization is over supporting Cartan subalgebras.
In the generic case, there is only one supporting Cartan subalgebra.
Nevertheless it would be nice if the minimization was redundant.

\begin{Problem}
\label{prob:cindep}
Is $S((|P_\alpha\ket{\psi}|^2)_\alpha)$ as introduced above
independent of the choice of supporting Cartan subalgebra?
\end{Problem}

Note that $S_C(\ket{\psi})$ is zero for any $\ket{\psi}$ contained in
a weight space for some Cartan subalgebra of $\lie{h}$ and that in
general, such states are not coherent for $\lie{h}$.  Furthermore,
these weight spaces are usually not one-dimensional.  Nevertheless,
this measure generalizes the bipartite setting.

\begin{Theorem}
\label{thm:supp_cartan}
Assume the bipartite setting with $\lie{h}=\lie{h}_l$.
The weight spaces of a supporting Cartan subalgebra for $\ket{\psi}$
are the one-dimensional spaces associated with tensor products
of Schmidt basis elements for each side for some choice of
Schmidt basis. Hence $S_C(\ket{\psi})=S(\ket{\psi})$.
\end{Theorem}

This implies that for the bipartite setting,
the answer to Problem~\ref{prob:cindep} is ``yes''.

\proof The projection of $\ketbra{\psi}{\psi}$ into $\lie{h}_l$ is
given by $\varrho=\rho_a\tensor\id/N_b+\id/N_a\tensor\rho_b -
\id/N_a\tensor\id/N_b$ where $\rho_a$ and $\rho_b$ are the respective
reduced density matrices. The supporting Cartan subalgebras are the
Cartan subalgebras that commute with $\varrho$.  These are necessarily
of the form $\lie{c}_a\tensor\id+\id\tensor\lie{c}_b$, where
$\lie{c}_a$ and $\lie{c}_b$ are $\cstar$-closed Cartan subalgebras of
$\lie{h}_a$ and $\lie{h}_b$ that commute with $\rho_a$ and $\rho_b$,
respectively.  Therefore, $\lie{c}_a$ ($\lie{c}_b$) is generated by the
projectors onto an orthogonal basis $B_a$ ($B_b$) of eigenstates of
$\rho_a$ ($\rho_b$, respectively). The associated weight spaces are
one-dimensional, spanned by tensor products of members of $B_a$ and
$B_b$.  Because the members of $B_a$ and $B_b$ can be paired to form a
Schmidt basis for $\ket{\psi}$, the result follows.  \qed

\section{The Convex Cones Setting}
\label{sect:convex}

Many of the notions introduced for $\cstar$-closed operator Lie
algebras can be generalized even further.  For example, we can work
with any linear space of operators and study properties of the convex
set of linear functionals induced by states.  In fact, as pointed out
in Section~\ref{sect:further}, there are physically interesting cases
where this may be necessary.  In this section we focus on the
convexity properties of the state space and investigate the extent to
which local maps and measures of generalized entanglement can
still be defined and retain their features.

\subsection{Convex Cones}

A \emph{convex cone} $C$ is a subset of a real linear space $U$ closed
under positive linear combinations.  That is, if $x,y\in C$ and
$p,q\geq 0$, then $px+qy\in C$. To avoid degeneracies, we assume that $U$
is the span of $C$. Let $\dot C$ consist of the non-zero elements of
$C$. The cone $C$ is \emph{pointed} if there is a linear functional
$\trace$ (the \emph{trace}) on $U$ such $\trace(\dot
C)>0$. Equivalently, $C$ is pointed if $C\cap (-C) = \{0\}$. We assume
that $U$ is finite dimensional and that $C$ is closed in the usual
topology for $U$. For the remainder of this paper, a \emph{cone} is a
closed, pointed, convex cone equipped with the positive linear
functional $\trace$.  For our purposes cones represent spaces of
unnormalized pure and mixed states.  In the Lie-algebraic setting, the
cone is given by the set of linear functionals $\lambda\in\lie{h}^*$
that are nonnegative multiples of $\lie{h}$-states.  The trace is
given by evaluation of $\lambda$ at the identity $\id\in\lie{h}$. If
$\lambda$ is induced by the matrix $\rho$, evaluation at the identity
gives the usual trace, $\trace(\rho)$.  We refer to members $x\in C$ with
$\trace(x)=1$ as \emph{states}.  The \emph{pure} states of $C$ are
extremal states of $C$.  Our assumptions on $C$ imply that every state
of $C$ is a convex combination of pure states.

In the Lie-algebraic setting, we explicitly introduced a second Lie
algebra $\lie{g}$ when discussing measures of relative entanglement.
Before we introduced such measures, $\lie{g}$ was implicitly present,
but was trivially associated with the set of all operators. This is
because the fact that $\lie{h}$-states are induced by density
operators plays a crucial role. In the convex cones setting, there is
no equally obvious way in which states are induced, so we explicitly
introduce an \emph{outer} cone $D\subseteq V$, whose states induce the
states on $C$ via a linear map $\pi:V\rightarrow U$ satisfying
$\pi(D)=C$, and $x\in D, \trace(x)=1$ implies $\trace(\pi(x))=1$, that
is, $\pi$ is \emph{trace preserving}.  In the Lie algebraic setting,
$\pi$ is simply the restriction map: If $\lambda$ is a
$\lie{g}$-state, then $\pi(\lambda)=\lambda\restrict\lie{h}\in
\lie{h}^+$.  We refer to $C$ as the \emph{inner} cone. If $x$ is a
pure state of $C$ then $\pi^{-1}(x)$ is convex closed and its extremal
states are pure states in $D$. 
\ignore{

To see this: If $\pi(y)=x$ and $y=\sum_k p_k y_k$, so
$x=\sum_k p_k \pi(y_k)$. Due to extremality
of $x$, $\pi(y_k) = x$. Hence $y$ is a convex combination
of elements of $\pi^{-1}(x)$.

}%
Note that in the Lie-algebraic setting, $\pi^{-1}(x)$ for a pure
$\lie{h}$-state $x$ is a pure $\lie{g}$-state.  We define
\emph{separable} states of $D$ to be states in the convex closure of
$\{\pi^{-1}(x)\suchthat \textrm{$x$ is pure in $C$}\}$. We denote the
cone generated by the separable states of $D$ as $D_{\textrm{\tiny
sep}}$ (this depends on $C$).  A pure state $x$ of $D_{\textrm{\tiny
sep}}$ satisfies that $\pi(x)$ is pure in $C$.

As we discuss the extent to which we can define suitable
generalizations of various notions to the convex cones setting, it is
worth keeping in mind what the two cones correspond to in the
bipartite setting. In this setting, $D$ is isomorphic to the cone of
positive operators on $\cH_a\tensor\cH_b$, with $\trace$ the usual
trace functional. The trace one operators are the density
matrices. $C$ is determined by the reduced density matrices. Formally,
$C$ is isomorphic to the cone of operators of the form $A\tensor
\id/N_b + \id/N_a \tensor B + \alpha\id/N_a\tensor\id/N_b$ with $A$,
$B$ traceless and $A+\alpha\id/N_a$ and $B+\alpha\id/N_b$ positive.
The connection to $\lie{h}_l$-states is discussed in
Section~\ref{sect:bi_lqo}.  The map from $D$ to $C$ takes $\rho$ to
$\trace_b(\rho)\tensor \id/N_b+\id/N_a\tensor\trace_a(\rho)
- \trace(\rho)(\id/N_a\tensor\id/N_b)$.

\subsection{Local Maps}

A \emph{positive} map of $D$ is a linear map ${A}:V\rightarrow V$ such
that ${A}(D)\subseteq D$. The map ${A}$ is \emph{trace preserving} if
$\trace(x)=\trace({A}(x))$ for all $x$. This definition corresponds to
positive, but not necessarily completely positive maps in the Lie
algebraic setting.  Without the algebraic structure available for
states, it is not possible to define a unique ``tensor product'' of
cones, as would be required to distinguish between positive and
completely positive maps~\cite{Namioka69a,Wittstock74a} (cited
in~\cite{Wilce92b}).  Because of the absence of a suitable tensor
product construction, we also do not have any suggestions for how to
address asymptotic questions by resource scaling.

\ignore{ There are two tensor products of cones that make sense.  The
first is the convex closure of the formal product of the cones'
vectors. The second is to use the dual space of the first construction
applied to the duals of the cones.  I think that these are the two
extremes, and the question is whether one can pick out something in
between. The only case where it looks like that may make sense without
adding the relevant algebraic structure is if the cones are self-dual
with respect to non-degenerate inner products on the real vector
spaces. In this case, one should pick a self-dual cone between the two
constructions. }%

The family of positive maps of $D$ is closed under positive
combinations and hence form a cone (without a trace). In the
Lie-algebraic, or even the bipartite setting, the extreme points of
this cone are not easy to characterize (see,
for example,~\cite{Wilce92b}, p. 1927,~\cite{Gurvits2002a}).  However, the
extreme points of the cone of completely positive maps are certainly
extremality preserving in the following sense: A positive map ${A}$ of
$D$ is \emph{extremality preserving} if for all extremal $x\in D$,
${A}(x)$ is extremal. There are extremality preserving positive, not
completely positive, maps. An example is partial transposition for
density operators of qubits. We call a positive map that is a mixture
of extremality preserving maps \emph{q-positive}. It is possible to
recapture the idea of complete positivity by explicitly introducing a
cone representing the ``tensor product'' extension of $D$.  This will
be discussed after defining liftability.  In the bipartite setting,
the family of q-positive maps of $D$ is between the family of positive
maps and the family of completely positive maps acting on density
matrices on $\cH_{ab}$.

The next step is to define a family of maps that generalizes the
separable maps.  Call a positive map ${A}$ of $D$ $C$-separable if it
is a mixture of extremality-preserving positive maps ${A}_k$ that are
also extremality-preserving and positive for $D_{\textrm{\tiny sep}}$.
In the bipartite setting, this definition includes maps such as the
swap, which exchanges the two subsystems and is not separable, in
addition to some non-completely positive operations. Note that if the
Lie-algebraic definition of separability is used, operations like the
swap are excluded because they are not in the Lie group generated by
$\lie{h}_l$: The swap induces an exterior automorphism of
$\lie{h}_l$. From the point of view of entanglement, including the
swap can make sense because it obviously does not increase
entanglement.

One tool used to narrow the family of separable quantum maps was based
on liftability. The definition of liftability immediately generalizes
to our cones.  We say that a positive map ${A}$ on $D$ can be
lifted to $C$ if ${A}$ preserves the nullspace of $\pi$, or,
equivalently, if there exists a positive map ${A}'$ on $C$ such
that $\pi({A}(x))={A}'(\pi(x))$. In this case, we say
that ${A}'$ is the lifting of ${A}$ to $C$.

Using liftability, we can add more cones to try to capture the idea of
complete positivity or to exclude maps like the swap.  For complete
positivity, introduce one more cone $E$ and positive trace-preserving
map $\sigma:E\rightarrow D$ (onto). In the setting where states are
defined by density matrices on a Hilbert space $\cH$ of dimension $d$,
$E$ represents the cone generated by density matrices on
$\cH\tensor\cH'$, with $\cH'$ of dimension at least $d^2$. With this
cone in hand, we can try to get the completely positive maps by
considering only maps that are a mixture of extremality preserving
maps ${A}_k$ obtained as liftings of extremality preserving
positive maps ${B}_k$ on $E$. Whether this works depends on the
answer to the following problem:

\begin{Problem}
\label{prob:cone_lie}
Let ${A}$ be a positive map on operators of
$\cH\tensor\cH'$ with $\dim(\cH')\geq \dim(\cH)^2$. Suppose
that ${A}$ preserves the set of rank one operators
and that it lifts to a map ${A}'$ of operators on $\cH$.
Is ${A}'$ completely positive?
\end{Problem}

To exclude the swap, it suffices to introduce cones included in $C$ to
represent density matrices on $\cH_a$ and $\cH_b$ and require
liftability to both of these cones.

The other tool used to restrict separable maps involves operators with
maximal ground spaces.  It is not clear how to apply
this tool to the convex cone setting since the distinction between
positive and negative eigenvalues is not easily recovered in the
action $\rho\rightarrow A\rho A^\cstar$.

To be able to generate families of maps by a kind of locality
preserving composition requires the idea of conditional composition
based on explicit maps.  An explicit positive map $\mathbf{A}$ on $D$
is given by $\mathbf{A}=(A_k)$ with $A_k$ extremality preserving
positive maps. For explicit separability, the $A_k$ are required to be
$C$-separable. In addition, we can impose the liftability condition on
each $A_k$. We call the latter \emph{explicit $C$-liftable separable
maps.}  The idea of Section~\ref{sect:li_lqo} to restrict the
separable maps by using certain minimal explicit separable maps can be
applied in the convex cones setting.  However, without the strong
symmetry present in the Lie-algebraic setting, the definition of
$k_{\textrm{\tiny min}}$ (Section~\ref{sect:li_lqo}) is unlikely to be
as natural. However, one could investigate the families of maps
obtained by replacing $k_{\textrm{\tiny min}}$ by $2,3,\ldots$.

Conditional composition can be used to generate a family of maps as
before.  One can then readily generalize communication complexity to
the resulting conditionally composed maps.

\subsection{Measures of Relative Entanglement}

The entanglement measures defined on the basis of a Schur-concave
function $S$ are intrinsically defined using only convexity.
Thus, for states $x\in C$,
\begin{equation}
  S(x) = \inf\{ S(\mathbf{p})\suchthat x=\sum_k p_k x_k\;\textrm{with $x_k$ pure}\},
\end{equation}
and for states $x\in D$,
\begin{equation}
S(x;C)=
\inf\{\sum_k p_k S(\pi(x_k)) \suchthat
       x=\sum_k p_k x_k\;\textrm{with $x_k$ pure}\}.
\end{equation}
In general, $S(x)$ is not concave, though this is the case in the
bipartite setting and if the set of states is a simplex. In the latter
case, the expression of a point as a convex combination of extreme
points is unique.

\begin{Problem}
For which convex sets is $S(x)$ concave for all Schur-concave $S$?
\end{Problem}

\begin{Theorem}
$S(x;C)$ is convex in $x$.
\end{Theorem}

\proof
Let $y=px_1+(1-p)x_2$ be a convex combination of states $x_1,x_2\in D$.
We show that $S(y;C)\leq p S(x_1;C)+(1-p)S(x_2;C)$, from
which the theorem follows.
For every way of expressing $x_k=\sum_l p_{kl} x_{kl}$
as a convex combination of pure states of $D$ we have
$y=\sum_l (p p_{1l}x_{1l}+ (1-p)p_{2l}x_{2l})$.
Thus
\begin{eqnarray*}
\begin{array}{rcll}
S(y;C)&\leq& \strutlike{\Big\{}\sum_l (p
      p_{1l}S(\pi(x_{1l}))+ (1-p)p_{2l}S(\pi(x_{2l})))
      & \textrm{by definition,}\\ 
   &=& \strutlike{\Big\{}p \sum_l
      p_{1l}S(\pi(x_{1l})) + (1-p)\sum_l p_{2l} S(\pi(x_{2l})).
\end{array}
\end{eqnarray*}
The last two sums can be chosen to be arbitrarily close
to $S(x_1;C)$ and $S(x_2;C)$.
\qed

Purity as defined in the Lie algebraic setting does not generalize to
the setting of convex cones unless $C$ has a well-defined center and
satisfies that all its pure states are equidistant from the center in
a natural metric.

\subsection{Monotonicity for Explicit Liftable Maps}

A desirable property for measures of entanglement is that they are
nonincreasing under the family of maps that are considered to be local.

\begin{Problem}
For which of the families of maps that we have introduced is $S(x;C)$
(or, more specifically, $S(x;\lie{h})$) nonincreasing? 
\end{Problem}

In the bipartite setting, it has been shown that $S(x;\lie{h}_l)$ is
nonincreasing under LOCC maps~\cite{Vidal2000b}.  Here we show that
this is the case in the convex cones setting for the family of
trace-preserving explicit liftable $C$-separable maps of cones. With
the cones that arise in the bipartite setting, this family of maps
includes the explicit liftable separable quantum maps. (See also
Problem~\ref{prob:cone_lie}.)  The monotonicity result is easy to see
for the later family because in this case, the family of maps consists
of mixtures of product unitaries.

For $x\not=0$ in a cone, define $\widehat
x=\widehat{\iboxlike{x}}(x)=\trace(x)^{-1} x$ to be the unique state proportional
to $x$. If $x=0$, define $\widehat x = 0$.  We say that the function
$\Upsilon:D\rightarrow\rls$ is \emph{explicitly nonincreasing} for the
trace-preserving explicit positive map $\mathbf{A}=(A_k)_k$ if for
extremal states $x\in D$,
\begin{equation}
\label{eq:vidal_cond}
\Upsilon(x)\geq\sum_k p_k \Upsilon(\widehat{A_k(x)}),
\end{equation}
where $p_k=\trace(\widehat{A_k(x)})$. 
The property of being explicitly nonincreasing
is useful as a sufficient condition for being nonincreasing.

\begin{Lemma}
\label{lemma:monolemma}
Suppose that $S(x;C)$ is explicitly
nonincreasing for the trace-preserving explicit positive map $\mathbf{A}$.
Then $S(x;C)$ is nonincreasing for $\mathbf{A}$.
\end{Lemma}

The Lemma holds for any $\Upsilon$ defined from its values on
pure states according to
$\Upsilon(x) = \inf\{\sum_k p_k \Upsilon(x_k)\suchthat
 x=\sum_k p_k x_k\;\textrm{with $x_k$ pure}\}$.

\proof Let $\mathbf{A}=(A_k)_k$ with $A_k$ positive and write
$p_k=\trace(A_k(x))$ To prove the lemma, first consider an
extremal $x$.  Then
\begin{eqnarray*}
\begin{array}{rcll}
S(\mathbf{A}(x);C) \strutlike{\Big\{}&=&
    S(\sum_k A_k(x);C) &\\
\strutlike{\Big\{}  &\leq&
   \sum_k p_k S(\widehat{A_k(x)};C) &\textrm{by convexity,} \\
\strutlike{\Big\{}  &\leq& S(x;C) &\textrm{by being explicitly nonincreasing.}
\end{array}
\end{eqnarray*}
For a nonextremal $x$, write $x=\sum_l q_l x_l$ with $x_l$ pure and
$\sum_l q_l S(x_l;C)$ arbitrarily close to $S(x;C)$. Note
that for pure $y$, $S(y;C)=S(\pi(y))$.
Then
\begin{eqnarray*}
\begin{array}{rcll}
S(\mathbf{A}(x);C) \strutlike{\Big\{} &=&
    S(\sum_l q_l\mathbf{A}(x_l);C) &\textrm{by linearity,}\\
\strutlike{\Big\{}  &\leq& \sum_l q_l S(\mathbf{A}(x_l);C) &\textrm{by convexity and trace preservation,}\\
\strutlike{\Big\{}  &\leq& \sum_l q_l S(x_l;C) &\textrm{by extremality of $x_l$.}
\end{array}
\end{eqnarray*}
The result now follows because the the right hand side is arbitrarily
close to $S(x,C)$.
\qed

\begin{Theorem}
\label{thm:monotonicity}
If $\mathbf{A}$ is a trace-preserving explicit liftable 
$C$-separable map of $D$, then $S(x;C)$
is explicitly nonincreasing under $\mathbf{A}$.
\end{Theorem}

\proof Let $\mathbf{A}=(A_k)_k$ with each $A_k$ liftable to $C$ and
$C$-separable.  Write $p_k=\trace(A_k(x))$.  Because of
Lemma~\ref{lemma:monolemma}, it is sufficient to prove
Inequality~\ref{eq:vidal_cond}. Let $x$ be a pure state of $D$. Let
$\pi(x)=\sum_l q_l y_l$ be a convex representation of $\pi(x)$ in
terms of pure states of $C$ such that $S(\mathbf{q})$ is arbitrarily
close to $S(x;C)=S(\pi(x))$.  We can find pure states $z_l\in D$ such
that $\pi(z_l)=y_l$.  Thus $x=\sum_l z_l + z$ for some $z$ with
$\pi(z)=0$.  With the appropriate interpretation of $A_k(x)/p_k$ when
$p_k=0$,
\begin{eqnarray*}
\begin{array}{rcll}
\pi(\widehat{A_k(x)}) 
\strutlike{\Big\{}  &=&
    \pi(A_k(x)/p_k) & \\
\strutlike{\Big\{}  &=&\pi(A_k(\sum_l q_l z_l + z)/p_k) & \\
\strutlike{\Big\{}  &=&
    \sum_l (q_l/p_k)\pi(A_k(z_l))  & 
       \textrm{since $A_k$ preserves the nullspace of $\pi$,} \\
\strutlike{\Big\{}  &=&
    \sum_l (r_{lk} q_l/p_k)\pi(\widehat{A_k(z_l)}) &
       \textrm{with $r_{lk} =\trace(A_k(z_l))$.}
\end{array}
\end{eqnarray*}
Since $A_k$ is $C$-separable and $z_l$ is pure in $D_{\textrm{\tiny
sep}}$, so is $A_k(z_l)$. Thus, by definition, $S(\widehat{A_k(x)};C)
= S(\pi(\widehat{A_k(x)}))\leq S((r_{lk} q_l/p_k)_l)$.  To prove the
desired inequality, bound as follows:
\begin{eqnarray*}
\begin{array}{rcll}
\sum_k p_k S(\widehat{A_k(x)};C) 
\strutlike{\Big\{}  &\leq& \sum_k p_k S((r_{lk}q_l/p_k)_l) & \\
\strutlike{\Big\{}  &\leq& S(\sum_k p_k(r_{lk}q_l/p_k)_l) & \textrm{by Schur concavity,} \\
\strutlike{\Big\{}  &=& S((\sum_k r_{lk} q_l)_l) & \\
\strutlike{\Big\{}  &=& S((q_l)_l) = S(\mathbf{q}) &\textrm{because $\mathbf{A}$ is trace preserving,}
\end{array}
\end{eqnarray*}
which is arbitrarily close to $S(x;C)$.
\qed

Conditional composition of trace-preserving explicit liftable
$C$-separable maps preserves explicit liftability and $C$-separability.
Nevertheless it is useful to know circumstances that guarantee
that conditional composition preserves monotonicity of $S(x;C)$.

\begin{Theorem}
Suppose that $S(x;C)$ is explicitly nonincreasing under the 
trace-preserving explicit
extremality-preserving maps $\mathbf{A}=(A_k)_k$ and
$\mathbf{B}_k$. Then it is explicitly nonincreasing under the
conditional composition $\mathbf{E}$ of $\mathbf{A}$ followed by the
$\mathbf{B}_k$. $\mathbf{E}$  is also an explicit extremality-preserving map.
\end{Theorem}

\proof Let $x$ be a pure state of $D$. That $\mathbf{E}$ is also an
explicit extremality-preserving map is clear.  Write $p_k =
\trace(A_k(x))$ and $q_{kl}=\trace(B_{kl}(A_k(x)))/p_k$. If $p_k=0$,
set $q_{kl}=0$. To prove Inequality~\ref{eq:vidal_cond}, compute
\begin{eqnarray*}
\begin{array}{rcll}
\sum_{kl} q_{kl} p_k S(\;\widehat{\iboxlike{x}}({B_{kl}(A_k(x))});C)\hspace*{-.5in}\\
  \strutlike{\Big\{}  &=&
    \sum_{kl} q_{kl} p_k S(\;\widehat{\iboxlike{x}}({B_{kl}(\widehat{A_k(x)})});C) & \\
  \strutlike{\Big\{}  &\leq&
    \sum_k p_k S(\widehat{A_k(x)};C) & 
      \textrm{because the $\mathbf{B}_k$ are explicitly non-} \\
      &&&\textrm{increasing and the $A_k(x)$ are extremal.} \\
  \strutlike{\Big\{}  &\leq&
    S(x;C) & \textrm{because $\mathbf{A}$ is explicitly nonincreasing.}
\end{array}
\end{eqnarray*}
\qed

\section{Discussion}

\subsection{Further Examples and Extensions}
\label{sect:further}

The traditional setting for studies of entanglement is that of
bipartite systems. Our investigation shows that the more general
theory based on Lie algebras exhibits most of the features associated
with bipartite entanglement, and a significant number of these
features can even be found in the convex cones setting.  As a result,
we hope that the general theory provides new insights into bipartite
entanglement and its generalizations to multipartite
systems. Relativizing the idea of entanglement has the advantage of
being able to immediately use the entire hierarchy of local Lie
algebras and associated entanglement measures in the multipartite
setting.

There are other settings where multiple, physically motivated Lie
algebras occur. We give four examples of such settings.  The first
example involves spectrum generating algebras (SGAs).  SGAs are used
to determine the spectrum (eigenvalues and eigenspaces) of quantum
systems. SGAs provide the starting point for one or more chains of Lie
subalgebras that are used for obtaining algebraic bases of states and
for expanding the Hamiltonian as a linear combination of invariant
(Casimir) operators belonging to the chains. When such an expansion
contains only invariant operators of a single algebraic chain, the
system exhibits a dynamical symmetry, and the corresponding spectrum
can be calculated exactly using the representation theory of Lie
subalgebras. In the generic case where operators from multiple chains
occur (that is, distinct dynamical symmetries coexist), the SGA
approach may still make it possible to accurately represent the
Hamiltonian in terms of a small number of algebraic operators.  Since
they were introduced in nuclear physics~\cite{Iachello1987a}, SGA
methods have been successfully applied to a variety of problems in
molecular, atomic, and condensed matter physics~\cite{Barut88a}.
Using the approach developed here, one could investigate the states'
relationships to the families of coherent states associated with the
Lie subalgebras and quantify their relative entanglement.

An example we have already mentioned as motivation for our work
involves fermions in $N$ modes. In this case, in addition to the
algebra of all relevant operators, there is the Lie algebra
$\lie{h}_p$ of number-preserving operators quadratic in the creation
and annihilation operators. These operators can be expressed in the
form $\mathbf{a}^\dagger M \mathbf{a}$ where $M$ is an $N\times N$
matrix. The $\lie{h}_p$-coherent states are the Slater
determinants (see, for example, \cite{Blaizot86a}, p.~7) and represent
independent fermions.  If the Lie algebra is enlarged to $\lie{h}_a$
consisting of all operators that are homogeneous quadratic in the
creation and annihilation operators, coherent states include BCS
states~\cite{Schrieffer64a}, which can be thought of as describing independent
fermion-pairs. Therefore, from this perspective, BCS states are unentangled.
On the other hand, they have entanglement with respect
to the pair $\lie{h}_p\subset\lie{h}_a$ of Lie algebras.

The example of fermions generalizes to anyons. Anyons as defined in
quantum field theory include particles with fractional exchange
statistics~\cite{Batista2002b}.  To apply our theory to
anyons requires using features of the convex cones setting.  This is
because the various sets of operators quadratic in the creation and
annihilation operators are Lie algebras only for fermions and
bosons~\cite{Lerda93a}.  This was one of our motivations for extending the
formalism.  The convex cones can be defined as the set of linear
functionals induced by states on sets of operators as before and
investigated using essentially the same basic tools. Further
investigation is required to determine whether special properties not
available in the convex cones setting still apply to quadratic anyonic
operator families.

For bosons in $N$ modes, four algebras frequently play an important
role.  The smallest, $\lie{h}_{pl}$ consists of the operators of the
form $\mathbf{a}^\dagger M \mathbf{a}$, where $M$ is an $N\times N$
matrix and $\mathbf{a}$ is the vector of annihilation operators of the
$N$ modes.  This algebra generates the passive linear optics
operators. A second Lie algebra, $\lie{h}_s$ is the one that generates
shifts in the canonical variables associated with the modes and
consists of operators at most linear in the creation and annihilation
operators.  The Lie algebra $\lie{h}_{al}\supseteq
\lie{h}_{pl}+\lie{h}_s$ consisting of all operators that are at most
quadratic in the annihilation and creation operators is the algebra that
generates all linear optics operators.  Finally, there is the algebra
of all relevant operators. The usual coherent states of optics and
harmonic oscillators are the $\lie{h}_{s}$-coherent states.

Although much of our proposal can be applied to the example of bosons,
caution is required in generalizing the finite dimensional theory to
the infinite dimensional state spaces of bosonic modes.  In addition,
algebras like $\lie{h}_s$ are not semisimple or reductive, requiring an
extension of the theory, as can be done for the theory of coherent
states~\cite{Perelomov85a,Zhang90a}.

\subsection{Relevance to Condensed Matter Physics}

Entanglement, and our generalizations of it, may be important in the
understanding of physical phenomena.  For example, the concept of
``quantum phase transitions'' \cite{Sachdev2001a} involves a
qualitative change in the behavior of correlation functions at zero
temperature, i.e. in a pure ground state, as parameters in a system's
Hamiltonian are varied.  In some cases an order parameter is
associated to the transition, in others a topological order.  Since
classical pure states cannot exhibit correlations, this is an
essentially quantum phenomenon. Moreover, the presence of correlations
between subsystems in a pure state can serve as a definition of
entanglement, so quantum phase transitions might be considered to be
due to a qualitative change in the nature of entanglement.  Therefore,
quantifying and classifying entanglement may help
characterizing a quantum phase transition.  Can measures of
entanglement distinguish between a broken symmetry and a topological
phase transition?  Can one classify quantum critical points?  It is
essential in this regard to have a notion of entanglement that need
not make reference to locality or subsystems.  Whether the correlation
functions that best characterize a given phase transition are those of
distinguishable subsystems (say, lattice sites) or some other kind of
correlations (say, two-particle correlation functions for systems of
indistinguishable particles) may determine whether standard
entanglement, or instead some generalization of it, provides
appropriate concepts.  Even standard entanglement is relative to a
distinguished factorization of a total Hilbert space into ``local''
ones, though this is usually unproblematic in quantum information
settings.  In other settings, such as many-body condensed-matter
systems, different factorizations may occur on a more equal footing
as ``global'' transformations typically play a natural role.
Thus a system of interacting bosons or fermions on a lattice may be
viewed in terms of a factorization of the state space into
distinguishable lattice sites, but the Fourier transformation from
position modes to momentum modes may provide an alternative
factorization; and it may also be that for some problems, correlations
between particles, rather than modes, are relevant, taking us beyond
the distinguishable-subsystems framework of standard entanglement
theory.

The introduction of ``quasiparticles'', or transformations such as the
Jordan-Wigner transformation~\cite{Jordan28a,Batista2001a}, may
further alter the algebraic language we use to analyze the system; our
motivation for such transformations may be mathematical (easier
solvability in one algebraic language than in another) or physical
(one algebra better exhibits the physical structure of the system's
dynamics, or of our interactions with it).  In either case, the
coherent states formalism is often known to be useful, and tools and
concepts from quantum information theory, such as generalized
entanglement measures, generalized LOCC and asymptotics may help as
well. Initial work in the direction of connecting the information
theoretic approach to entanglement to condensed matter can be found
in~\cite{Osborne2001a,Osborne2002a,Osterloh2002a}.

To give a more explicit example, Landau quasiparticles refer to those
{\it dressed} particles of the original interacting system that weakly
interact as a result of {\it transferring} most of the real
interactions into the properties of the quasiparticles themselves. As
a result, these quasiparticles may be qualitatively different from the
original particles, an example of which is provided by the composite
fermions in the quantum Hall setup \cite{Jain98a}. But how do we
construct those quasiparticles?  Weak interactions can be related to
weak correlations and, therefore, {\it weak generalized entanglement}.
If one can re-express the original problem in a language such that
the Hamiltonian operator belongs to the quadratic expressions in the
language's generating operators (for example, creation and annihilation
operators) then we know that the quasiparticles are
non-interacting. Otherwise, we need to quantify the degree of
``entanglement'' (in the ground state, say) to determine whether the
{\it particles} generated by the language interact sufficiently weakly
to behave as true quasiparticles.  The use of hierarchical languages
may help to address this issue \cite{Batista2002a}.

\subsection{Conclusion}

We have outlined a program whose goal is to tie together the theory of
entanglement and the theory of coherent states.  We implemented the
first few steps of this program starting with the observation that
fundamental concepts of the theory of coherent states coincide with
concepts from the theory of entanglement. We extended this observation
by providing general definitions of the key information-theoretic
notions in entanglement theory. In particular, we introduced several
classes of quantum maps to the Lie algebraic setting appropriate
for coherent state theory that generalize the idea of separable maps
for multipartite systems and approach LOCC. The numerous open problems
attest to the richness of this program. 

After noting that many of the notions that we generalized can, to some
extent, be stated even more generally in the context of convex cones,
we made this explicit by investigating appropriate definitions for
convex cones. Except for the convex cones arising as spaces of linear
functionals on operator families induced by states, most such convex
cones are not physically relevant. Nevertheless, they help us
appreciate what aspects of the various models are required in order to
investigate different properties of generalized entanglement and their
information-theoretic implications.

The main conclusion of our program so far is that conventional
entanglement is a special case of a much more general theory with many
of the same features. Furthermore, it is clear that entanglement is a
relative property of states, requiring that states that are mixed from one
perspective can be pure from other, more powerful perspectives.
Once this relativity is recognized, it is possible to investigate
relative entanglement of states when many physically motivated
perspectives coexist. Examples include multipartite systems, condensed
matter systems, and systems whose dynamics is described by the chain
of Lie algebras associated with a dynamical symmetry or a spectrum
generating algebra.

\begin{appendix}
\section{Comparison of the Settings for Generalized Entanglement}
\label{table:comparison}

The following table shows the three settings as generalizations of the
bipartite setting. 

\pagebreak

\newcommand{\srrght}{\small\raggedright\setlength{\baselineskip}{12pt}}

\vspace*{-.5in}
\begin{center}
\hspace*{-.4in}\begin{tabular}{|p{1in}|p{1.4in}|p{2.3in}|p{2.3in}c@{}|}
\hline
  & \fhf Bipartite systems\fhf 
  & \fhf Lie algebras\fhf 
  & \fhf Cones\fhf &
\\\hline\hline
\srrght Structure: 
  & \srrght 
    $\cH_a\tensor \cH_b$, a tensor product of two Hilbert spaces.
  & \srrght 
    $\{\id\}\subseteq\lie{h}\subseteq\lie{g}$,
    $\cstar$-closed Lie algebras of operators on a Hilbert space $\cH$.
  & \srrght 
    Closed, convex cones $C\subseteq D$ with traces,
    and $\pi:D\rightarrow C$ a linear, trace-preserving, map onto $C$.
  &
\\\hline
\srrght States:
  & \srrght
    Full or reduced density matrices.
  & \srrght
    Linear functionals on $\lie{h}$ or $\lie{g}$
    induced by density matrices.
  & \srrght
    Trace-one elements of $C$ or $D$.
  &  
\\\hline
\srrght Specialization to bipartite systems: 
  &
  & \srrght 
    $\lie{h} = \{A\tensor\id+\id\tensor B\}$,
    $\lie{g}$ is the set of all operators on $\cH_a\tensor\cH_b$.
  & \srrght 
    $C$ $\simeq$ $\{$ $(A,B)$ $\suchthat$ $A$ ($B$) positive on $\cH_a$ ($\cH_b$)$\}$,
    $D$ $\simeq$ $\{C$ $\suchthat$ $C$ positive on $\cH\}$,
    $\pi(C)=(\trace_a(C),\trace_b(C))$.
  &
\\\hline
\srrght Specialization to Lie algebras:
  &
  &
  & \srrght
    $C$ ($D$) consist of the linear functionals induced on
    $\lie{h}$ ($\lie{g}$) by
    positive $\rho$ on $\cH$ as
    $x\rightarrow \trace(\rho x)$. $\pi$ is the restriction map.
  &
\\\hline
\srrght Distinguished pure states: 
  & \srrght 
    Product pure states. 
  & \srrght 
    Coherent (or, equivalently, pure) $\lie{h}$-states.
  & \srrght 
    States $x \in D$  such that $\pi(x)$ is pure in $C$.
  &
\\\hline
\srrght Distinguished mixed states:
  & \srrght 
    Separable states.
  & \srrght 
    Convex combinations of $\lie{g}$-states that
    restrict to coherent $\lie{h}$-states. 
  & \srrght 
    The cone $D_{\textrm{\tiny sep}}$ of separable states in $D$ consisting of
    convex combinations of states $x\in D$ such that $\pi(x)$ is pure
    in $C$.
  &
\\\hline
\srrght Pure state entanglement measures:
  & \srrght 
    Von-Neumann entropy for pure states.\par
    Unilateral purity. 
  & \srrght 
    $S$ Schur concave, $\lambda$ an $\lie{h}$-state:
    $S(\lambda)$ $=$
    $\inf\{S(\mathbf{p})$ $\suchthat$ $\lambda=
          \sum_k p_k\lambda_k$ with $\lambda_k$ $\lie{h}$-coherent, 
          $p_k \geq 0\}$.\par
    $\lie{h}$-purity.\par
    Measures based on amplitudes ($S_a(\lambda)$) 
    and supporting Cartan subalgebras ($S_C(\lambda)$).
  & \srrght 
    For $x$ a pure state in $C$,
    $S(x)$ $=$
    $\inf\{S(\mathbf{p})$ $\suchthat$ $x$ $=$
          $\sum_k p_k x_k$ with $x_k$ pure, $p_k\geq 0 \}$.
  &
\\\hline
\srrght Mixed state entanglement measures:
  & \srrght
    Given pure state entanglement measure $S$:
    $S(\rho)$ $=$ $\inf\{ \sum_k p_k S(\rho_k)$
    $\suchthat$ $\sum_k p_k\rho_k = \rho$, $\rho_k$ is
    a pure product state, $p_k\geq 0 \}$.
  & \srrght
    Given an $\lie{h}$-state measure $S$ and a $\lie{g}$-state $\lambda$,
    $S(\lambda)$ $=$ $\inf\{ \sum_k p_k S(\lambda_k)$
    $\suchthat$ $\sum_k p_k\lambda_k = \lambda$, $\lambda_k\restrict\lie{h}$ is
    coherent, $p_k\geq 0 \}$.
  & \srrght
    Given a $C$-measure $S$, $x$ a state in $D$.
    $S(x)$ $=$ $\inf\{ \sum_k p_k S(\pi(x_k))$
    $\suchthat$ $\sum_k p_k x_k = x$, $\pi(x_k)$ is
    pure, $p_k\geq 0 \}$.
  &
\\\hline
\srrght Properties of entanglement measures:
  & \srrght
   Convex. Monotone under LOCC.
  & \srrght
   Convex. Monotone under explicit liftable separable quantum maps.
  & \srrght
   Convex. Monotone under trace-preserving
   explicit liftable $C$-separable maps of $D$.
  &
\\\hline
\srrght Maximally entangled states:
  & \srrght
    Bell states.
  & \srrght
    See~\cite{Klyachko2002a}.
  & \srrght
    Undefined.
  &
\\\hline
\srrght Non-classicality of entangled states:
  & \srrght
    Bell inequalities.
  & \srrght
    See~\cite{Klyachko2002a}.
  & \srrght
    Undefined.
  &
\\\hline
\srrght Hierarchies:
  & \srrght 
    Add the unilateral algebras.
  & \srrght 
    Arbitrary family of operator Lie algebras ordered by inclusion.
  & \srrght 
    Arbitrary family of cones, partially ordered by trace-preserving
    onto maps. &
\\\hline
\end{tabular}
\end{center}

\pagebreak

\vspace*{-.5in}
\begin{center}
\hspace*{-.4in}\begin{tabular}{|p{1in}|p{1.4in}|p{2.3in}|p{2.3in}c@{}|}
\hline
& \fhf Bipartite systems\fhf 
  & \fhf Lie algebras\fhf 
  & \fhf Cones\fhf &
\\\hline\hline
\srrght Local unitary operators:
  & \srrght 
    Product unitary operators. 
  & \srrght 
    $e^{i\real(\lie{h})}$.
  & \srrght 
    Positive linear isomorphism $f:D\rightarrow D$ such that
    $\pi f = \tilde f \pi$ for some isomorphism $\tilde f: C\rightarrow C$. 
    \par
    Caution: Defs of local maps do not
    always specialize to the corresponding defs for Lie algebras.
  &
\\\hline
\srrght Local operators:
  & \srrght 
    Product operators. 
  & \srrght 
    $\strutlike{\Big\{}\overline{e^{\lie{h}}}$.
  & \srrght 
    $C$-product maps:
    Extremality-preserving positive maps $f:D\rightarrow D$
    that preserve extremality in $D_{\textrm{\tiny sep}}$ also.
  &
\\\hline
\srrght Separable maps:
 & \srrght
   $\rho\rightarrow \sum_k A_k \rho A_k^\dagger$, where the
   $A_k$ are product operators.
 & \srrght
   $\rho\rightarrow \sum_k A_k\rho A_k^\dagger$, where 
   $\strutlike{\Big\{}A_k\in\overline{e^{\lie{h}}}$.
   \par
   Caution: Defs of local maps do not always specialize to
   the corresponding defs for bipartite systems.
 & \srrght
   $C$-separable maps:
   $x\rightarrow\sum_k A_k(x)$, where the $A_k$ 
   are $C$-product maps
 &
\\\hline
\srrght Unilocal operators:
  & \srrght $
    A\tensor I$, $I\tensor A$.
  & \srrght 
    Operators of $\lie{h}$ with maximal ground spaces?
    Operators whose action lifts to $\lie{h}$-states?
  & \srrght 
    $C$-product maps of $D$ that lift to $C$?
  &
\\\hline
\srrght Compatible families of one-sided local operators:
  & \srrght 
    Operators acting on the same subsystem.
  & \srrght 
    Operators conjugate under $e^{\lie{h}}$ to one
    with maximal ground spaces?
  & \srrght 
    Undefined.
  &
\\\hline
\srrght LOCC:
  & \srrght 
    Monoid generated by conditional composition
    of explicit unilocal quantum maps.
  & \srrght 
    Monoid generated by conditional composition
    of explicit quantum maps consisting of compatible families?
    \par
    Monoid generated by conditional composition
    of explicit liftable separable quantum maps?
    \par
    \ldots
  & \srrght 
    Monoid generated by conditional composition of
    trace-preserving explicit liftable $C$-separable maps?
  &
\\\hline
\srrght Communic\-ation complexity:
  &
  \multicolumn{3}{c}{\small Defined in terms of outcome probabilities
   in each step of a conditional composition.}
  &
\\\hline
\srrght Known monotonicity of entanglement results:
  & \srrght
    Under LOCC maps.
  & \srrght
    Under explicitly liftable separable quantum maps.
  & \srrght
    Under trace-preserving explicit liftable $C$-separable maps.
  &
\\\hline
\srrght Resource scaling:
  & \srrght 
    By tensor product, preserving orientation of the bipartition.
  & \srrght 
    Grow Lie algebras over tensor products of $\cH$ using
    partial traces. May require additional structure?
  & \srrght 
    Undefined.
  &
\\\hline
\end{tabular}
\end{center}

\end{appendix}

\pagebreak

\acknowledgments We thank Cristian Batista for the original motivation for
investigating entanglement in a subsystem-independent way by pointing
out that Slater determinants should be considered unentangled.  We
also thank the other ``coherent tangles,'' in particular Jim
Gubernatis, Leonid Gurvits, Rolando Somma and Wojciech Zurek for
stimulating discussions. We acknowledge support from the DOE (contract
W-7405-ENG-36). 

\bibliography{cs}

\end{document}